\documentclass[useAMS,usenatbib]{mn2e}

\usepackage{epsfig, natbib, graphicx, color, amsmath, cancel,icomma,bm, amssymb}

\input epsf 

\usepackage{color}

\def\lesssim{\mathrel{\hbox{\rlap{\hbox{\lower3pt\hbox{$\sim$}}}\hbox{\raise2pt\hbox{$<$}}}}}
\def\gtrsim{\mathrel{\hbox{\rlap{\hbox{\lower3pt\hbox{$\sim$}}}\hbox{\raise2pt\hbox{$>$}}}}}

\newcommand{\Mpc}{\mbox{Mpc}}

\newcommand{\msun}{M_\odot}

\newcommand{\avg}[1]{\left\langle #1 \right\rangle}

\newcommand{\be}{\begin{equation}}
\newcommand{\ee}{\end{equation}}
\newcommand{\bea}{\begin{eqnarray}}
\newcommand{\eea}{\end{eqnarray}}

\newcommand{\kpc}{\mbox{kpc}}

\newcommand{\Planck}{{\it Planck}}
\newcommand{\Euclid}{{\it Euclid}}

\newcommand{\redmapper}{redMaPPer}

\newcommand{\zspec}{z_{\mathrm{spec}}}

\newcommand{\photoz}{photo-$z$}

\newcommand{\PSZ}{PSZ1}

\title[redMaPPer III: Comparison with the Planck 2013 Cluster Catalog]{redMaPPer III: A Detailed Comparison of the Planck 2013 and SDSS DR8 RedMaPPer Cluster Catalogs}

\author[E. Rozo, E. S. Rykoff, J. G. Bartlett, J.B. Melin]{E. Rozo$^{1}$, E. S. Rykoff$^1$, James G. Bartlett$^{2,3}$, Jean-Baptiste Melin$^4$.\\
$^{1}${SLAC National Accelerator Laboratory, Menlo Park, CA 94025, U.S.A.}\\
$^{2}${APC -- Universit\'{e} Paris Diderot, CNRS/IN2P3, CEA/lrfu, Observatoire de Paris, Sorbonne Paris Cit\'{e}, 75205 Paris Cedex 13, France.}\\
$^3${Jet Propulsion Laboratory, California Institute of Technology, Pasadena, California, U.S.A.}\\
$^4${DSM/Irfu/SPP, CEA-Saclay, F-91191 Gif-sur-Yvette Cedex, France.}
}

\begin{document}

\maketitle

\label{firstpage}

\begin{abstract}
We compare the \Planck\ Sunyaev-Zeldovich (SZ) cluster 
sample (\PSZ) to the Sloan Digital Sky Survey (SDSS)  \redmapper\ catalog, finding that all \Planck\ 
clusters within the \redmapper\ mask and within the redshift range probed by \redmapper\ 
are contained in the \redmapper\ cluster catalog.  These common clusters define
a tight scaling relation in the richness-SZ mass ($\lambda$--$M_{SZ}$) plane, with an intrinsic scatter in richness of 
$\sigma_{\lambda|M_{SZ}} = 0.266 \pm 0.017$.  The corresponding intrinsic scatter in true cluster halo mass at fixed
richness is $\approx 21\%$.   The regularity of this scaling relation is used to identify failures in
both the \redmapper\ and \Planck\ cluster catalogs.  Of the 245 galaxy clusters in common, we identify
three failures in \redmapper\ and 36 failures in the \PSZ.  Of these, at least 12 are due to clusters whose optical counterpart
was correctly identified in the \PSZ, but where the quoted redshift for the optical counterpart in the external data base
used in the \PSZ\ was incorrect.
The failure rates for \redmapper\ and the \PSZ\ are $1.2\%$ and $14.7\%$ respectively, or 9.8\% in the \PSZ\ after subtracting the external
data base errors.  We have further identified
5 \PSZ\ sources that suffer from projection effects (multiple rich systems along the line-of-sight of the SZ detection) and
17 new high redshift ($z\gtrsim 0.6$) cluster candidates of varying degrees of confidence.  
Should all of the high-redshift cluster candidates identified here be confirmed, 
we will have tripled the number of high redshift \Planck\ clusters in the SDSS region.
Our results highlight the power of multi-wavelength observations to identify and characterize systematic
errors in galaxy cluster data sets, and clearly establish photometric data both as a robust cluster finding method,
and as an important part of defining clean galaxy cluster samples.  
\end{abstract}


\section{Introduction}

The abundance of galaxy clusters as a function of mass is a well known cosmological 
probe \citep[][and many others]{henryetal09,oukbir1992,bartlett1993,eke1996,viana1996,vikhlininetal09b,rozoetal10a,mantzetal10a}.   
As such, it is of critical
importance for galaxy cluster surveys to control the level of systematic failures in the cluster selection
and redshift assignment, lest the corresponding cosmological inferences be biased.

One way to test a particular cluster selection algorithm is to utilize multi-wavelength data to establish
the scaling relations of galaxy clusters: gross deviations from the mean behavior 
can be used to flag systems that may be subject to otherwise unsuspected systematic effects.
In \citet[][hereafter Paper II]{rozorykoff13} we performed an extended analysis of this type in order to characterize the failure
rate in the recently published Sloan Digital Sky Survey (SDSS) \redmapper\ cluster catalog \citep[][hereafter Paper I]{rykoffetal13}.
In this context, ``failure'' refers to either false cluster identifications or 
incorrect cluster property assignments, e.g., redshifts and/or richness. 
In that work, we estimated the \redmapper\ failure rate at $\sim 1\%$.

Paper II was limited partly by the availability of X-ray data at higher redshifts,  partly by the large
statistical noise in how X-ray luminosity traces cluster mass, and mostly by the lack of high-resolution
X-ray data for complete optical samples.   Consequently, \Planck\ data on a large number of galaxy clusters
provides additional critical consistency tests to help better characterize failures in the \redmapper\ cluster 
finding algorithm.   
We note that while in Paper II we did in fact consider \Planck\ Sunyaev-Zeldovich (SZ) data in our analysis, at the time only the 
early \Planck\ SZ results had been published, which did not allow for the significantly more
detailed analysis performed in the present work.
Here, we extend the analysis of Paper II to include the newly released \Planck\ SZ cluster sample \citep[\PSZ,][hereafter PXXIX]{PXXIX}.

Our analysis also tests the robustness of the \PSZ\ cluster selection algorithm, complementing the validation 
presented in PXXIX with the well-characterized \redmapper\  optical catalog.  While PXXIX encountered 
difficulty in using their selected optical samples to uniquely identify \PSZ\ sources, we 
demonstrate that this difficulty is not endemic to photometric cluster catalogs.
Specifically, the \redmapper\ galaxy cluster richness produces a well defined, tight scaling relation
with SZ mass proxies, and outliers of this relation always signal a systematic error in either the \redmapper\
or the \PSZ\ cluster catalogs.  Our results firmly establish optical data as a critical component 
of multi-wavelength efforts aimed at defining well controlled, fully characterized galaxy cluster samples.

Our paper is laid out as follows.   Section~\ref{sec:data} introduces the various data sets employed in our study.
Section~\ref{sec:results} describes our analysis and presents our results, and Section~\ref{sec:conclusions}
presents our conclusions.  Throughout, we adopt a fiducial flat $\Lambda$CDM cosmology with $\Omega_m=0.3$
and $h=0.7$, consistent with the choice in PXXIX.


\section{Data}
\label{sec:data}

\subsection{The Planck XXIX Cluster Catalog}

The \Planck\ satellite \citep{tauber2010} was launched on May 14, 2009, as the third generation space mission dedicated 
to cosmic microwave background observations.  The \Planck\ payload consists of two instruments, the Low Frequency 
Instrument covering 30, 44 and 70~GHz \citep{bersanelli2010, mennella2011}, 
and the High Frequency Instrument with bands centered at 100, 143, 217, 353, 545, and 857~GHz 
\citep{lamarre2010, PlanckHFI2011}, with angular resolution varying from 33\arcmin to 5\arcmin.  
A series of early results were published in 2011 \citep{planck11_mission}, and the first cosmology results were 
published in 2013 along with 15.5 months of science data \citep{planck2013-I}.

The \PSZ\ is based on the first 15.5 months of \Planck\ observations \citep{PXXIX}.  The SZ detection uses the six highest \Planck\
frequency channels, spanning 100\,GHz to 857\,GHz.  Prior to detecting SZ sources, a Galactic mask and a point source mask based
on the Planck Catalogue of Compact Sources \citep{pXXVIII} are applied.  The resulting
holes in the maps are filled in prior to searching for SZ sources, and the catalog is trimmed around the mask.
SZ sources are detected using three different types of algorithms: 2 matched filter algorithms \citep[MMF1, MMF3,][]{melinetal06}
and a Bayesian finder called Powell--Snakes \citep[PwS,][]{carvalhoetal12}.  All three algorithms place priors on the cluster spectral
and spatial characteristics, which are in turn employed to distinguish SZ sources from random noise fluctuations.
A comparison of various SZ cluster selection algorithms is presented in \citet{melinetal12}.

After running each of the three cluster finders, all $S/N>4.5$ sources are collated into a master catalog,
which is cleaned for obvious contamination based on the \Planck\ high frequency spectral information.
The SZ signal strength, measured by the integrated Compton-$y$ parameter $Y_{SZ}$, of the remaining 
sources is typically strongly degenerate with the cluster size, a problem
that is resolved by the use of informative priors on the relation between $Y_{SZ}$ and cluster size.  Since cluster
radii are typically defined in terms of matter overdensity criteria, such a prior takes the form of a fiducial
$M$--$Y_{SZ}$ relation, so that a by-product of the measurement is a cluster mass estimate for each system.
PXXIX utilizes the $M$--$Y_{SZ}$ relation of \citet{arnaudetal10} for these purposes, and a full description 
of the algorithm will appear in a future publication \citep[Arnaud et al., in preparation, see also][]{gruenetal13}.

In a detailed study, \citet{rozoetal12b,rozoetal12c,rozoetal12d} noted significant differences in published
cluster X-ray mass calibrations and argued that the mass calibration adopted in 
\citet{arnaudetal10} would lead to tension between observed cluster abundance and the cosmological 
constraints from the cosmic microwave background \citep{rozoetal13}.  This was borne out in the
first results from \Planck.  Consequently, we do {\it not} take the ``SZ-mass'' ($M_{SZ}$) reported 
in PXXIX as an accurate determination of cluster total mass.  However, the reported SZ mass is a well defined SZ observable 
--- indeed, it is mathematically equivalent to $Y_{SZ}$ --- and thus it can be used to define scaling relations and to 
study the regularity of the \Planck\ and \redmapper\ cluster samples.  

Finally, we note that in order to estimate $Y_{SZ}$ and the corresponding SZ mass $M_{SZ}$, a cluster redshift is needed for each detection.
In the \PSZ, these redshifts are obtained either by cross matching with various catalogs, or through extensive photometric and spectroscopic
follow-up of \Planck\ SZ sources.  When matching to external catalogs, PXXIX follows a rank-ordered priority list.
The top priority is the MCXC catalog \citep{piffarettietal11}, followed by NED and SIMBAD, followed by redshifts
from the \citet{wenetal12} SDSS photometric cluster catalog.   A few remaining clusters get their redshifts either from
additional external photometric catalogs, or from an internal analysis of SDSS data within the \Planck\ team.  
Many clusters (more than 200) were also subjects of both photometric and spectroscopic follow-up.  The final 
PXXIX SZ cluster catalog is a compilation of all this data.

Throughout, we only consider the list of galaxy clusters labelled in PXXIX as confirmed systems.
We restrict ourselves to systems
with reported redshifts in the ranges $z \in [0.08,0.6]$ since
$z=0.6$ is the maximum
redshift at which \redmapper\ can detect massive clusters in the SDSS with reliable redshift estimates.
When considering scaling relations, 
we exclude clusters with $z>0.5$ since clusters above this
redshift have very large measurement errors in their richness.


\subsection{The SDSS DR8 \redmapper\ Cluster Catalog}

\redmapper\ is a new red-sequence photometric cluster finding algorithm which was recently applied
to the SDSS Data Release 8 \citep{dr8}.  The algorithm and SDSS DR8 catalog
is described in detail in Paper I.  A detailed comparison of \redmapper\ to other photometric
cluster finding algorithms is presented in Paper II, which also includes a multi-wavelength study of the
performance of \redmapper\ in the SDSS.

Briefly, \redmapper\ iteratively self-trains a model of red-sequence galaxies calibrated by
exploiting an initial seed spectroscopic galaxy sample.  The associated
spectroscopic requirements are minimal, and easily satisfied by existing SDSS spectroscopy.
Once the red-sequence model has been trained, the algorithm attempts to grow a galaxy
cluster centered about every SDSS photometric galaxy.  The galaxies are first rank-ordered
according to their likelihood of being a central galaxy.  Once a rich galaxy cluster
has been identified ($\lambda\geq 5$, where
$\lambda$ is the number of red-sequence galaxies hosted by the cluster), 
the algorithm iteratively determines a photometric redshift
based on the calibrated red-sequence model, and recenters the clusters about 
the best cluster center, as gauged from the photometric data.   
The final published cluster catalog is then further trimmed
to a richness limit of $\lambda \ge 20$ for $z\leq 0.35$.  Above this redshift, the catalog
becomes ``flux'' limited due to the SDSS survey depth, and the richness limit increases
rapidly with redshift.  Roughly speaking, richness measurements are reliable out to
$z=0.5$.  For $z\in [0.5,0.6]$, clusters can be detected, but their richness measurements
become very noisy.  When run on SDSS data, 
automated cluster finding is not really feasible with the \redmapper\ algorithm above redshift $z=0.6$.

In what follows, we will match the \Planck\ cluster catalog to the \redmapper\ cluster catalog.
One obvious concern of this type of exercise is that
a \Planck\ cluster may go unmatched if the system falls below the \redmapper\ detection threshold.
In order to minimize this possibility, we always match the \Planck\ catalog to our own private
\redmapper\ catalog, which lowers the detection threshold from $\lambda=20$ to $\lambda=5$.
Doing so increases the \redmapper\ cluster sample by almost a factor of 16, from
$\approx 26,000$ clusters to $\approx 412,000$ systems.  
As it turns out, however, all good \Planck--\redmapper\ cluster matches
result in pairs of richness $\lambda \ge 20$.  


\subsection{The MCXC Cluster Catalog}

The Meta-Catalogue of X-ray Clusters \citep[MCXC:][]{piffarettietal11} is a compilation of galaxy clusters based on
publicly available X-ray data from both the
ROSAT All Sky Survey~\citep[RASS:][]{vogesetal99} and serendipitous searches
in ROSAT pointed observations.  The RASS contributing catalogs are
NORAS~\citep{bohringeretal00}, REFLEX~\citep{bohringeretal04}, BCS~\citep{ebelingetal98}, 
SGP~\citep{cvbcr02}, NEP~\citep{hmvbb06}, MACS~\citep{eeh01}, and CIZA~\citep{emt02,kemt07},
while the contributing serendipitous catalogs are 160D~\citep{mmqvh03},
400D~\citep{bvheq07}, SHARC~\citep{rnhup00}, WARPS~\citep{hpejs08}, and
EMSS~\citep{gmsws90}.  The data from each of the individual galaxy catalogs was
collected and homogenized, deleting duplicate entries, and enforcing a
consistent X-ray luminosity definition.  In the catalog, $L_X$ is defined to be the
X-ray luminosity in the 0.1--2.4~keV band within an $R_{500c}$ aperture.  We will use this
catalog as a baseline for understanding the centering offset distribution between
the \Planck\ and \redmapper\ cluster centers.


\section{Analysis of \PSZ\ Clusters}
\label{sec:results}


\subsection{Cluster Matching}
\label{sec:matching}

We match the \PSZ\ to the SDSS DR8 \redmapper\ catalog
using a simple angular matching algorithm.   
To match the two catalogs, we rank order the \PSZ\ clusters by signal-to-noise (S/N).  Starting from the
top (largest S/N) cluster, we define its match as the richest \redmapper\ system within $10\arcmin$
of the reported \Planck\ detection.   If a match is found,
the corresponding \redmapper\ system is removed from the list of candidate \redmapper\ matches, and we move
on to the next \PSZ\ cluster.  We emphasize that our matching criteria does not impose any restrictions
on the redshifts of matched cluster pairs.  This is purposely so, as we wish to compare the redshifts
reported in the two cluster catalogs.   Of course, this also implies that all \Planck--\redmapper\ cluster
pairs should only be considered tentative associations, 
pending the results of our full analysis, summarized in Section \ref{sec:consolidation}.


\begin{figure}
\hspace{-5mm} \includegraphics[width=90mm]{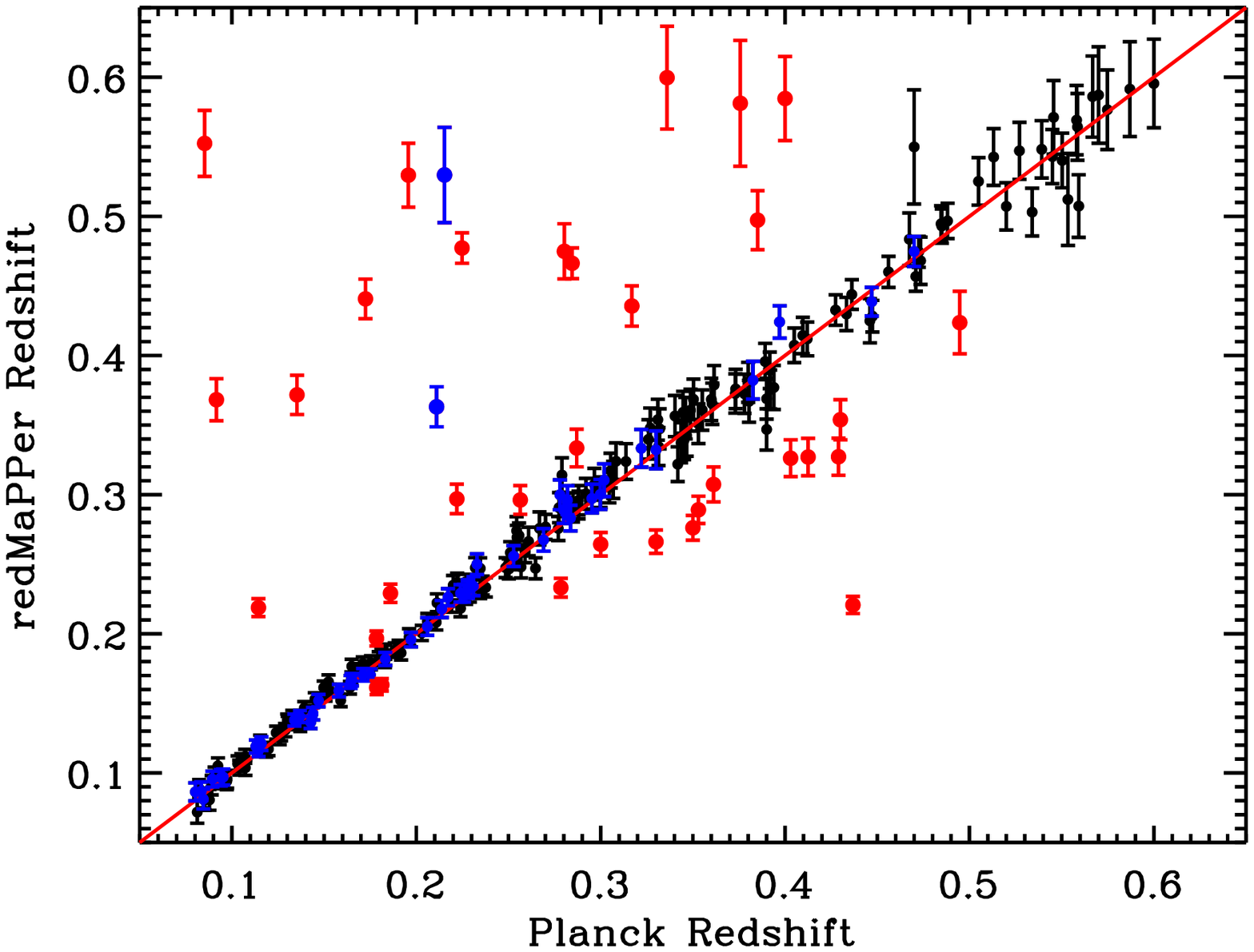}
\caption{Comparison of \PSZ\ redshift to the redshift of its tentative \redmapper\ match.  The matches will be revised based on the results
of our subsequent analysis.  Blue points are systems included in the PXX
cosmology analysis \citep{pXX}.  Black points are clusters for which the \redmapper\ and \PSZ\
redshifts are consistent with each other.  There are 33 $3\sigma$ outliers, shown in red,
and 2 additional obvious outliers in blue that form part of the PXX cosmology
sample.  We discuss the outliers in Section~\ref{sec:zout}.
} 
\label{fig:zcomp}
\end{figure}


Figure \ref{fig:zcomp} compares the \PSZ\ cluster redshifts to those of the
matched \redmapper\ clusters.  We restrict ourselves to the redshift range 
$ z\in[0.08,0.6]$, where the \redmapper\ redshifts are expected to be accurate.  There are
245 \PSZ\ systems in this redshift range within our angular mask, 35 of which show
up as redshift outliers.  These outliers will be discussed in Section \ref{sec:zout}.
For now, we will focus on the sub-sample of galaxy clusters where the
two redshift estimates agree.  This way, by characterizing the relation between optical
and SZ data first using our well matched clusters,  we can later on use this
information to resolve the redshift conflicts shown in Figure~\ref{fig:zcomp}.


\subsection{The $\bm{\lambda}$--$\bm{M_{SZ}}$ Scaling Relation: }
\label{sec:scaling}

\begin{figure}
\hspace{-5mm} \includegraphics[width=90mm]{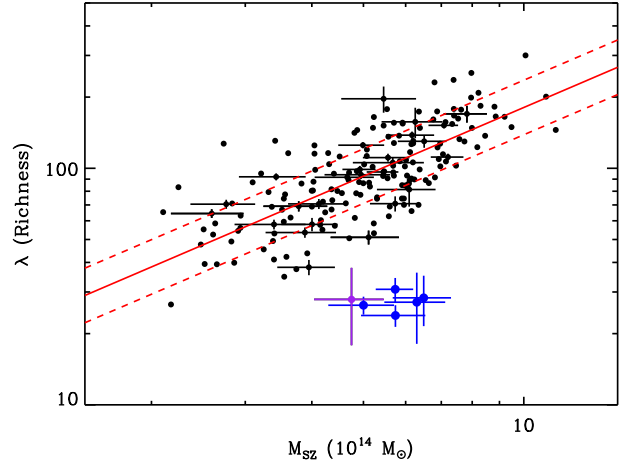}
\caption{Relation between cluster richness $\lambda$ and SZ-mass $M_{SZ}$.  
The cluster richness is evaluated at the optical center, and is taken directly from the \redmapper\ cluster catalog.
The SZ-mass is taken directly from PXXIX.  Only clusters with consistent redshift matches are included (i.e., black points
in Fig.~\ref{fig:zcomp}).
To avoid clusters with large richness errors, we restrict our analysis
to systems with $z_\lambda \leq 0.5$; this leaves a total of 191 objects.  
The red solid line is the best fit scaling relation,
while the dashed lines mark the $1\sigma$ intrinsic scatter.  Blue points are $3\sigma$ outliers. 
\Planck\ 299 is shown as a purple point with error bars.
Only a small random fraction of clusters is shown with error bars to avoid overcrowding the
plot. 
} 
\label{fig:scaling}
\end{figure}

We consider the $\lambda$--$M_{SZ}$ relation of the galaxy clusters for which the \PSZ\ and \redmapper\
redshifts agree.  Since the richness errors become very large at $z > 0.5$, we further restrict
our analysis to systems with \redmapper\ redshift $z_\lambda \leq 0.5$ (this is the redshift 
assigned by \redmapper\ photometry).  This results in a sample
of 191 galaxy clusters, shown in Figure~\ref{fig:scaling}.    We see that there is a tight
relation between the \redmapper\ cluster richness $\lambda$ and the SZ-mass $M_{SZ}$, though
the existence of a small outlier population with low richness and high SZ mass is immediately apparent.

We have fit this data using the Bayesian
fitter employed in \citet{rozoetal12a}, with an automated outlier rejection of $3\sigma$ outliers. 
We find
\bea
\avg{\ln \lambda|M_{SZ}} & = & a+\alpha\ln \left( \frac{M_{SZ}}{5.23\times 10^{14}\ \msun} \right)\\
a & = & 4.572 \pm 0.021 \\
\alpha & = & 0.965\pm 0.067 \\
\sigma_{\ln \lambda|M_{SZ}} & = & 0.266 \pm 0.017.
\eea
If we assume no intrinsic covariance between cluster richness and SZ signal at fixed mass, and an intrinsic scatter of the 
SZ-based mass estimates of $15\%$ (20\%), the corresponding estimate of the intrinsic scatter in true halo mass
at fixed richness is $\approx 21\%$ (17\%) \citep[see][for how the observed scatter $\sigma_{\ln \lambda|M_{SZ}}$
is related to the desired scatter $\sigma_{\ln M|\lambda}$]{rozoetal12c}.
These values are broadly consistent with those
derived in Paper II based on $M_{gas}$.  Our results are also robust to lowering the maximum
cluster redshift $ z \leq z_{max}=0.5$.


Our algorithm identifies five outliers.  The first of these is Abell 963 (\Planck\ 617),
which was identified in Paper II as having a systematically low richness because
of a small region of bad photometry around a bright star in SDSS.
For the remaining four outliers, we find no similar \redmapper\ failures,
suggesting that the SZ detection by \Planck\ is {\it not}
sourced by the \redmapper\ cluster, and therefore that the assigned \PSZ\ cluster
redshift is incorrect.  In all four cases, visual inspection
of the corresponding SDSS and WISE fields revealed galaxy overdensities with faint, very red
optical counterparts, suggesting all four are good high redshift ($z\geq 0.6$) cluster candidates.
These clusters are \Planck\ 1093, 719, 729, and 441.  The RA and DEC of the corresponding candidate galaxy overdensities
are reported in Table \ref{tab:unmatched}, which includes DR8 photometric or spectroscopic redshifts where available.

\begin{figure}
\hspace{-5mm} \includegraphics[width=90mm]{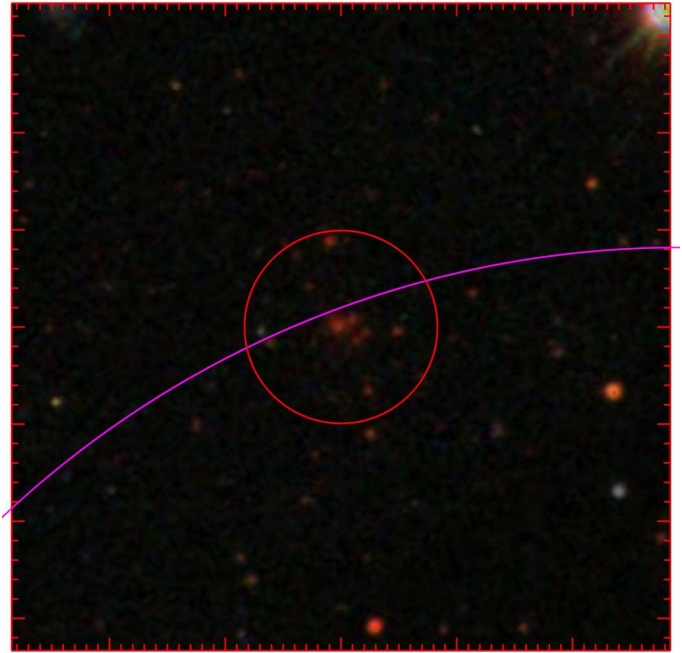}
\caption{Proposed high redshift cluster match to \Planck\ SZ source 299.  The cluster candidate (red circle)
shown above is at a distance of $3\arcmin$ from the reported \Planck\ location for the SZ source 299.
The purple circle segment has a radius of $3\arcmin$, and is centered at the reported \Planck\ location.
The entire box is $2\arcmin\times 2\arcmin$, centered on the optical high $z$ cluster candidate.
The central galaxy of the proposed cluster candidate has a spectroscopic redshift $\zspec=0.748$.
The red circle has a radius of $0.3\arcmin$ or $\approx 130\ \kpc$.
} 
\label{fig:299}
\end{figure}

We note that there is one additional system that appears to be an outlier by eye, \Planck\ 299.  The cluster
is not flagged as an outlier by our automated algorithm because of its large richness error bar.  The \redmapper\
match is located $5.5\arcmin$ away from the reported \PSZ\ location.  Visual inspection of the SDSS
field reveals a very obvious high redshift ($\zspec=0.748$) cluster match $3\arcmin$ away, located
at  $RA=231.6383$, $DEC=54.1520$, shown here in Figure~\ref{fig:299}. 
Thus, we consider our automated \redmapper\ association (and therefore the \PSZ\ redshift)
incorrect.  Of course,
a definitive statement would require confirmation of our proposed cluster match, and for said cluster
to obey either X-ray--SZ or optical--SZ scaling relations.  


Before we move on, we emphasize that this analysis does {\it not} properly account for selection effects,
so we caution against relying on the above scaling relation for precision work.  Our only goals in this work 
are to demonstrate the existence of a tight scaling relation, to highlight the regularity
of optical richnesses as a mass tracer, and to demonstrate the utility of optical data in understanding and improving
the \Planck\ SZ cluster catalog.

In short, of the 210 non-redshift outliers in Section \ref{sec:matching}, only 191 have $z\leq 0.5$.  These define
a tight scaling relation, from which we estimate a scatter in mass at fixed richness of $\approx 21\%$.  There are, however,
6 outliers.  One outlier is due to bad photometry in SDSS,
while the remaining five systems were incorrectly associated by our matching algorithm and thus have incorrect
redshifts in \PSZ.  Visual inspection reveals possible high redshift ($z\geq 0.6$) counterparts
for all five of these systems.   Interestingly, all 6 of these systems were SZ sources matched to low richness
clusters in the \citet{wenetal12}, \citet{haoetal10}, or \citet{szaboetal11} catalogs.


\subsection{Cluster Centering}
\label{sec:centering}

\begin{figure}
\hspace{-5mm} \includegraphics[width=90mm]{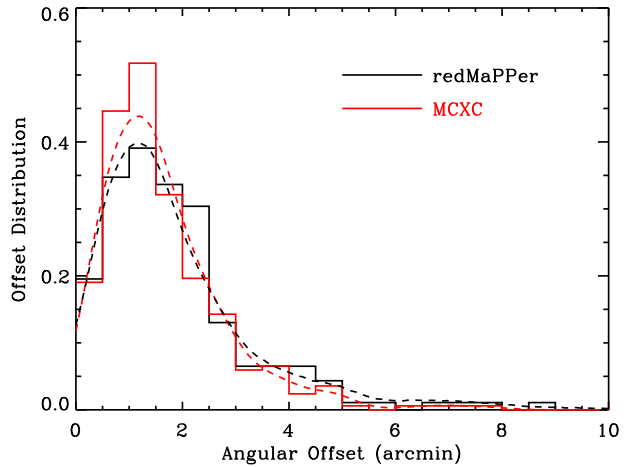}
\caption{Distribution of angular offset between \Planck\ location and the corresponding \redmapper\ or MCXC
cluster center, as labeled.  Both sets of clusters are restricted to the same redshift range.  The red and black dashed
curves are smooth models for each of the distributions (see text). 
} 
\label{fig:offsets}
\end{figure}

Figure \ref{fig:offsets} shows the distribution of angular offsets of the \PSZ\ clusters relative to the \redmapper\ centers.
As in the previous section, we restrict our analysis to galaxy clusters where the \redmapper\ and \PSZ\ redshifts agree.
Moreover, we remove from the sample the six galaxy clusters identified as erroneous cluster matches in Section \ref{sec:scaling}.
For comparison, we also show the corresponding
distribution for \Planck--MCXC cluster matches (red histogram).  The two offset distributions (\Planck--\redmapper\ and \Planck\--MCXC) 
are not consistent with each other,
as determined via a KS test.
We attribute this difference to miscentering in the optical.  To test this hypothesis, we first use a  Kernel Density Estimator (red dashed line)
to obtain a smooth model of the \Planck--MCXC centering distribution. 
This model is then convolved
with a simple miscentering model for \redmapper\ systems,  in which 85\% of the clusters are correctly centered, and the
remaining 15\% have a centering offset that is uniformly distributed out to $1\ \Mpc$.
The resulting distribution of angular offsets is the black dashed line.  A KS test reveals that the model
is consistent with the observed distribution for \Planck--\redmapper\ matches.
We note that the difference between the black and red dashed curves in Figure \ref{fig:offsets} is relatively minor: optical miscentering
acts to transfer a small but non-negligible amount of probability from the peak at $1\arcmin$ separation to a tail at $\approx 5\arcmin$
separation.


We have visually inspected all galaxy clusters where the offset relative to \redmapper\ is larger than $5\arcmin$.  
There are ten such clusters.  Of these, two clusters, \Planck\ 278 and 445, appear to be simple statistical fluctuations
in the centering offsets, which are $5.0\arcmin$ and $5.3\arcmin$ respectively. That is, the \Planck--\redmapper\ 
matches appear to be reasonable, and we do not find other structures within the field of view; the clusters simply
appear to populate the large centering offset tail of the distribution.
Four clusters (\Planck\ 728, 249, 472, and 587) are systems with two obvious galaxy clumps,
which not surprisingly can lead to large offsets when the SZ center falls closer to the opposite component from
that selected by \redmapper, but it is not entirely clear which component is dominant.  In this context,
we do not rely heavily on the \Planck\ centering since \Planck\ centers are themselves highly uncertain,
and in some cases our visual inspection suggests that the main cluster component is the one that is further
from the \Planck\ position \citep{PXXIX}.  A more definitive statement about these clusters will require high resolution X-ray data.
One cluster (\Planck\ 113) is an obvious case of optical miscentering based on visual inspection
of the field.

One cluster (\Planck\ 280) is matched to a rich \redmapper\ cluster ($\lambda=126.5$, $z_\lambda=0.324$),
but the match is $6.6\arcmin$ ($1.9\ \Mpc$) away.  We were able to identify a galaxy overdensity
in SDSS and WISE that may be a better match to this \Planck\ detection, at $RA=340.6030$,
$DEC=17.5214$, $z=0.86\pm 0.10$.  The quoted photoz is that of the putative central galaxy.
We expect our original tentative association 
is incorrect, and that the \Planck\ detection corresponds to a high redshift galaxy cluster.   We note the 
angular distance to \Planck\ 280 is $\approx 2\theta_{500}$, so it is also possible that the system
is suffering from projection effects.

Of the remaining two clusters, cluster \Planck\ 77 may or may not be a statistical fluctuation.  The centering
offset to its \redmapper\ match is very large, $7.7\arcmin$, and the optical center is obviously correct upon visual inspection.
However, \Planck\ 77 is a low redshift cluster ($z=0.118$), so the observed offset is in fact less than the assigned
$\theta_{500}$ angular radius.
We are also unable to identify galaxy overdensities in WISE that suggest a high redshift cluster.
Finally, we note, however that \Planck\ 77 is a low S/N system that is only detected by one of the three algorithms
used to construct the \PSZ\ sample \citep{PXXIX}.  Because of the large angular offset and the fact that the cluster is only detected by one of the three cluster finding algorithms, we consider \Planck\ 77 a possible false detection.

\begin{figure}
\hspace{-2mm} \includegraphics[width=87mm]{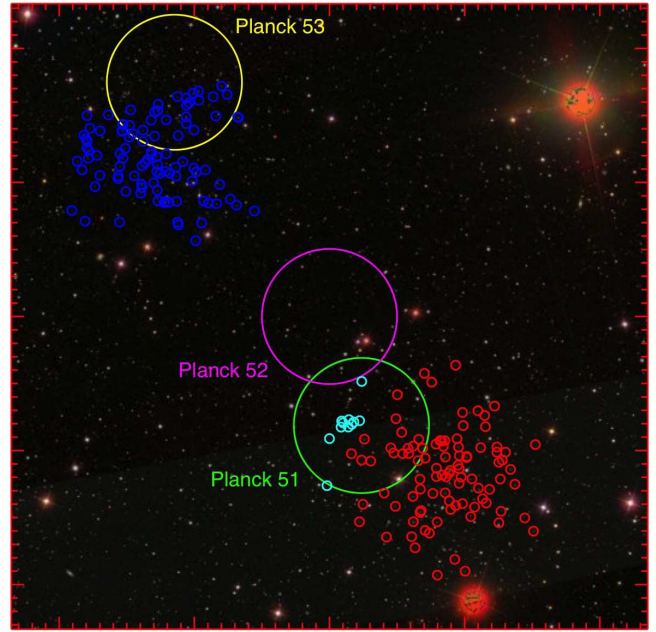}
\caption{SDSS image of the sky around \Planck\ SZ sources 51, 52, and 53.
Each \Planck\ cluster is marked with a $3\arcmin$ circle, as labeled.
There are 3 rich \redmapper\ clusters
in the region, for which we have circled galaxies brighter than $0.5L_*$
with a membership probability $p\geq 0.5$.  Each \redmapper\ cluster
is assigned a different color: RM 1995 is red ($z_{\rm spec}=0.338$), RM11143 is cyan($z_{\rm spec}=0.443$), and RM 689 is blue
($z_{\rm spec}=0.382$).
} 
\label{fig:mess}
\end{figure}

The remaining large centering offset cluster is \Planck\ 52, which PXXIX notes forms a complicated
triple system with \Planck\ 51 and \Planck\ 53.  With \redmapper\ we are able 
to shed considerable light on the region.  The entire field containing clusters
\Planck\ 51 through 53 is shown in Figure~\ref{fig:mess}.  
The southernmost rich \redmapper\ cluster is RM 1195 ($\lambda=85.4$, $\zspec=0.338$, member galaxies in red), 
which is matched to \Planck\ 51. 
Note, however, that \Planck\ 51 sits directly on top of RM 11143 ($\lambda=23.5$, $\zspec=0.443$,
member galaxies in cyan).
This suggests that \Planck\ 51 is picking up the SZ signal of the richer \redmapper\ cluster RM 1195, but that the
\Planck\ detection is miscentered due to a projection effect in the SZ between RM 1195 and RM 11143, pulling the SZ detection
towards the latter system.


The northernmost \Planck\ cluster is \Planck\ 53,
which is a good match to RM 689 ($\lambda=97.0$, $\zspec=0.382$, member
galaxies in blue).  Note that
the spectroscopic redshifts of \Planck\ 51 and \Planck\ 53 are significantly different ($\zspec=0.338$ vs $\zspec=0.382$),
corresponding to a line of sight distance of $\approx 180\ \Mpc$, or nearly $10^4\ \mathrm{km/s}$, suggesting these structures
are not correlated.  

This leaves us with \Planck\ 52, which sits on an essentially empty piece of sky, roughly halfway between
\Planck\ 51 and \Planck\ 53.  It has three candidate matches with
richness $\lambda \ge 20$, two of which are better associated with \Planck\ 51 (RM 1995 and RM 11143 discussed above),
and one of which is clearly associated with \Planck\ 53.  We have inspected both the SDSS and WISE images for a possible
high redshift cluster counter part, but were unable to find one.  

To explain the origin of \Planck\ 52, we recall that the \PSZ\ is the union of three distinct
cluster finding algorithms, two matched filter methods (MMF1 and MMF3) and a Baysian source detector called Powell--Snakes (PwS).
Clusters \Planck\ 51 and 53 were only detected by the PwS method, while \Planck\ 52 was only detected by the MMF3 method \citep{PXXIX}.  
This strongly suggests that the MMF3 detection is a blend of the \Planck\ 51 and \Planck\ 53 clusters, explaining why it sits
in between the two systems and has no good \redmapper\ association.  This is consistent with the large positional error assigned by 
MMF3 to this detection.
In short, we consider \Planck\ 51 as affected by projection effects in the SZ, \Planck\ 52 we flag as a false detection (really, a blend of \Planck\  51 and 52),
and \Planck\ 53 is robustly associated with RM 689.  

In short, of the ten clusters with large angular separations, two we consider good matches, four are double clusters, one is an obvious
case of optical miscentering, one is a high redshift cluster candidate, and two are labeled as false detections.


\subsection{Redshift Outliers}
\label{sec:zout}

We now turn our attention to the 35 redshift outliers in Figure~\ref{fig:zcomp}.  These outliers 
could arise in one of three ways.  If the \Planck\ and \redmapper\ clusters have been properly
matched, then either the PXXIX redshift is incorrect, or the \redmapper\ redshift is incorrect.
Alternatively, the matching algorithm may have failed.   We now determine
the origin of each of the redshift conflicts identified in Figure~\ref{fig:zcomp}.

\subsubsection{Visual Inspection}
\label{sec:visual}

To begin with, we check our cluster matching by visually inspecting the SDSS and WISE
images for each of the \PSZ\ clusters that resulted in a redshift conflict in Section \ref{sec:matching}.
Our goal is to ensure that every \PSZ\ cluster is matched to the best possible \redmapper\ match. 
We emphasize that ``best possible'' does not mean correct; we sometimes find \Planck\ sources
that have no convincing \redmapper\ match.  In such cases, we let the original match stand,
with the expectation that subsequent analysis will confirm those clusters as poor matches.
It should also be noted that ``unconvincing'' is a qualitative decision made by us based on the
visual inspection.  Quantitative tests will be presented in subsequent sections.
Our visual inspection also revealed several \Planck\ detections where the SZ signal is likely to
be affected by projection effects, i.e., there are multiple rich galaxy clusters along the line-of-sight 
of the \Planck\ detection.  These clusters are flagged as such.


\begin{table*}
\centering
\caption{Summary of our visual inspection of the 35 $3\sigma$ outliers in Fig.~\ref{fig:zcomp}.  
The ``Cluster'' column is the index used as the unique identifier in the \PSZ.  The next two columns compare the \PSZ\ redshift with 
the \redmapper\ redshift estimate $z_\lambda$.
When possible, the correct cluster redshift is written in {\bf bold}.  Spectroscopic redshifts in
the ``Comment'' column are taken directly from SDSS.}
\begin{tabular}{lccl}
\hline
Cluster & \PSZ-$z$ & $z_\lambda$ & Comment \\
\hline
     513 & 0.211 & \bf 0.363 $\pm$ 0.014  &  $\zspec=0.350$. \\
     1216 & \bf 0.215 & --- & Matching procedure failure due to \redmapper\ incompleteness. \\
     391 & 0.350 & \bf 0.275 $\pm$ 0.009 & DR8 \photoz: $z=0.265 \pm 0.015$. \\
     732 & 0.225 & --- & $\zspec=0.472$.  SZ-projection with cluster at $\zspec=0.225$. \\
   1128 & 0.085 & --- & $\zspec=0.559$. SZ-projection with cluster at $\zspec=0.088$. \\
     660 & 0.222 & \bf 0.293 $\pm$ 0.010  &  $\zspec=0.296$.  \\
     622 & 0.495 & --- & Unconvincing.  Possible high $z$ match,  $RA=42.8736$, $DEC=-7.9581$. \\  
      484 & 0.317 & --- & Unconvincing.  Possible high $z$ match,  $RA=178.1366$, $DEC=61.3629$, $z=0.86\pm 0.11$. \\    
     316 & 0.430 & \bf 0.352 $\pm$ 0.014  &  $\zspec=0.350$ \\
     537 & 0.353 & \bf 0.287 $\pm$ 0.010  &  $\zspec=0.284$ \\
      73 & 0.0916 & --- & SZ-projection of clusters at $\zspec=0.091$ and $\zspec=0.386$. \\
     888 & 0.412 & \bf 0.330 $\pm$ 0.014  &  $\zspec=0.323$ \\
      500 & \bf 0.280 & 0.514 $\pm$ 0.039  &  Bad SDSS photometry. \\
     222 & 0.181 & \bf 0.162 $\pm$ 0.005  &  $\zspec=0.163$ \\
     865 & 0.278 & \bf 0.234 $\pm$ 0.007  &  $\zspec=0.227$ \\
     303 & 0.300 & \bf 0.274 $\pm$ 0.009  &  $\zspec=0.269$ \\
      97 & 0.361 & \bf 0.310 &  Possible projection, but one candidate ruled out by scaling relation.\\
     668 & 0.287 & \bf 0.329 $\pm$ 0.013 &  $\zspec=0.335$ \\ 
     574 & 0.196 & --- & Unconvincing.  Possible high $z$ match,  $RA=170.9893$ and $DEC=43.0600$, $z=0.72\pm 0.13$. \\
     510 &  0.330 & --- & Projection of two clusters with no obvious best match.\\
     443 & 0.437 & \bf 0.221  &  Possible projection, but one candidate ruled out by scaling relation.\\
     764 & 0.285 & \bf 0.454 $\pm$ 0.011 &  $\zspec=0.462$ \\
     216 & \bf 0.336 & \bf 0.359 $\pm $ 0.016 & $\zspec=0.336$.  Original match incorrect, see Appendix. \\
    1123 & 0.114 & \bf 0.219 $\pm$ 0.006 & $\zspec=0.233$.  See text for discussion. \\
      308 & 0.178 & \bf 0.196 $\pm$ 0.005     &  DR8 \photoz: $0.200\pm 0.010$. \\
      768 & 0.403 & --- & Unconvincing.  Possible high $z$ match,  $RA=158.28934$, $DEC=13.79361$, $z=0.58\pm 0.13$. \\
     779 & 0.256 & \bf 0.294 $\pm$ 0.010 & $\zspec=0.299$. \\
      292 & 0.186 & \bf 0.229 $\pm$ 0.007 & $\zspec=0.221$.  \\
     234 & 0.400 & --- &  DR8 \photoz: $z=0.579 \pm 0.018$. SZ-projection with cluster at $z_\lambda=0.594\pm 0.035$. \\
     678 & 0.376 & ---  & Unconvincing.  Possible high $z$ match,  $RA=147.0918$, $DEC=24.7901$, $z=0.59\pm 0.04$. \\
     376 & 0.178 & ---      &  Unconvincing.  Possible SZ projection of clusters at $\zspec=0.159$ and $\zspec=0.178$. \\
     505 & 0.172 & --- & Unconvincing. Possible high $z$ match,  $RA=172.23425$, $DEC=59.94395$, $z=0.64\pm 0.08$.   \\   
     416 & 0.135 & \bf 0.373 $\pm$ 0.014 &  DR8 \photoz: $z=0.40\pm 0.02$. \\
     527 & 0.385 & --- & Unconvincing.  Possible high $z$ match,  $RA=21.71503$, $DEC=-7.20770$, $z=0.72\pm 0.04$.  \\   
      13 & 0.429 & \bf 0.325 $\pm$ 0.013 &  $\zspec=0.312$ \\
\hline
\end{tabular}
\label{tab:zout}
\end{table*}


Our results are summarized in Table~\ref{tab:zout}.  The table includes both the \PSZ\ redshift, and the
\redmapper\ photometric redshift of the 
best \redmapper\ match as determined from visual inspection.  When a match is ambiguous, we do 
not report a \redmapper\ redshift.  In the cases where a correct cluster redshift can be unambiguously identified,
the corresponding redshift is written in {\bf bold}. 
In brief, we find
\begin{itemize}
\item 3 clusters  where we confirm that the \PSZ\ redshift is correct (\Planck\ 1216, 500, and 216).  The original redshift conflicts were due to
	two incorrect cluster associations (including one incorrect match due to \redmapper\ incompleteness), and one incorrect \redmapper\
	redshift due to bad SDSS photometry.
\item 19 clusters where we confirm that the original association with \redmapper\ clusters is correct, and the \redmapper\ redshift is correct.
\item 8 unconvincing cluster matches.  For these clusters, we expect the \PSZ\ redshift to be incorrect, and for our best \redmapper\ match
	to prove unsatisfactory as well.  In 7/8 cases, our visual inspection reveals a candidate high redshift match for the \Planck\ detection,
	with varying degrees of confidence.
\item 5 clusters where the SZ signal appears to be sourced by an SZ-projection effect. These are \Planck\ 732, 1128, 73, 510, and 376.
\end{itemize}


\begin{figure}
\hspace{-2mm} \includegraphics[width=87mm]{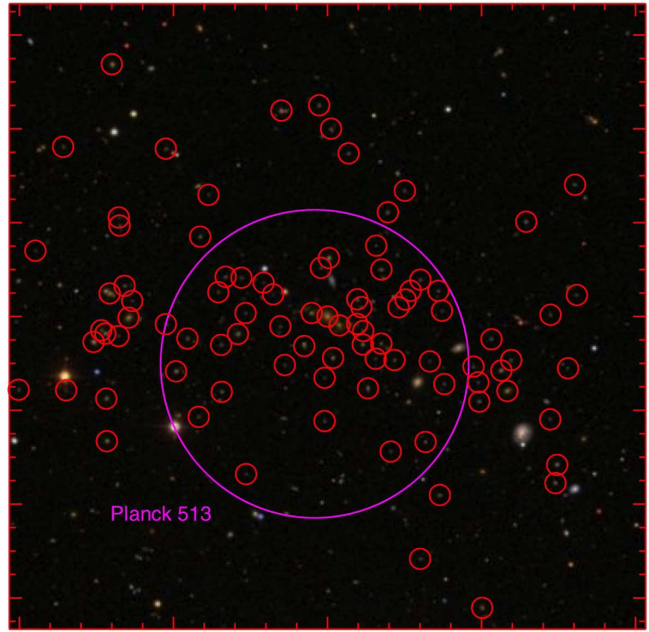}
\caption{SDSS image for \Planck\ 513.  The large purple circle shown is 
$3\arcmin$ ($\approx 0.9\ \Mpc$)
in radius, centered on the \Planck\ location.  There is an obvious cluster of galaxies at this
location, with spectroscopic redshift $\zspec=0.350$ (RM 87).  
However, this cluster
was assigned a redshift of $z=0.211$ in the \PSZ.  
The red circles are 
galaxies with \redmapper\ membership probability $p\geq 0.8$ brighter than $0.8L_*$.
} 
\label{fig:513}
\end{figure}


Comments on individual systems can be found in Appendix \ref{app:comments}.
Here, we showcase only two systems.  The first is cluster \Planck\ 513, which is the
only \PSZ\ cluster in the SDSS region that belongs to the \Planck\ cosmology
sample in \citet{pXX} and has been assigned an incorrect redshift.
The SDSS image of this SZ source is shown in Figure~\ref{fig:513}.  There is
an obvious cluster in the image (RM 87), which itself has an obvious central galaxy.
The central galaxy has a spectroscopic redshift available, from which we see
$\zspec=0.350$.  There are also multiple spectroscopic members that confirm this
redshift.  By contrast, the assigned redshift in PXXIX is $z=0.211$.  This redshift
was matched to Abell 1430, and the association is correct, but the cluster
redshift taken from SIMBAD is clearly erroneous.
We have
chosen to highlight this cluster because it formed part of the cosmology sample
in \citet{pXX}, not because it is a particularly egregious example of an incorrect redshift:
there were multiple cluster for which the correct cluster redshift was equally obvious.

\begin{figure}
\hspace{-2mm} \includegraphics[width=87mm]{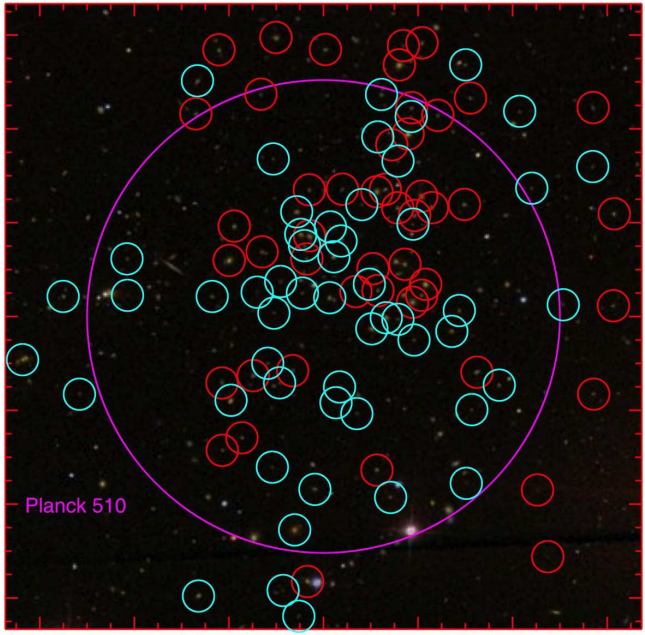}
\caption{SDSS image for \Planck\ 510.  The large purple circle shown is 
$3\arcmin$ in radius, centered on the \Planck\ location.  There are 2 \redmapper\ clusters
at this location, RM 128 and RM 141, each with at 5 spectroscopic
members that confirm the \redmapper\ cluster redshift for their parent cluster.
Cluster members with membership probability $p\geq p_{0.8}$ (see text) and luminosity
$L\geq 0.8L_*$ are shown with red (RM 128) and cyan (RM 141) circles.
Evidently, \Planck\ 510 is a spectacular example of an SZ projection effect.
}
\label{fig:510}
\end{figure}

The second system we would like to highlight is \Planck\ 510, which is a spectacular
example of an SZ projection.  The SDSS image centered
on the \Planck\ detection is shown in Figure~\ref{fig:510}.  The purple circle is $3\arcmin$
in radius, and is centered at the \Planck\ location.  There are two nearby rich
\redmapper\ clusters, RM 128 and RM 141.  We have circled all cluster members
with membership probability $p\geq p_{0.8}$ and luminosity $L\geq 0.8L_*$ for each
of these clusters in red and cyan, respectively.  Here, $p_{0.8}$ is the probability
threshold accounting for 80\% of the membership probability of the clusters, i.e.,
\be
0.8\lambda = \sum_{p\geq p_{0.8}} p_i.
\ee
We use this criterion for showing cluster galaxies to ensure comparable membership selection
between the low and high redshift clusters.\footnote{Had we used a fixed
probability cut (say $p\geq 0.8$), we would have selected more members for the low redshift cluster than
for the high redshift cluster because membership assignment is  less secure at high redshifts.}
By applying a probability threshold that is based on a fixed fraction of the cluster richness,
we ensure that comparable fractions of cluster members are selected.


We see from Figure~\ref{fig:510} that the two \redmapper\ clusters are obviously
collocated in the sky, and both coincide with the \Planck\ SZ detection.
Both systems have 5 members with spectroscopic redshifts, from which
we derive cluster redshifts $\zspec=0.2566$ and $\zspec=0.3715$.  
These compare well 
with the \redmapper\ photometric
redshift estimates $z_\lambda = 0.267 \pm 0.008$ and $z_\lambda=0.372\pm 0.014$.
In short, there is  no doubt that these clusters are two separate systems projected along the line
of sight, and that the deprojection by \redmapper\ via photometric data was successful.
The richnesses of the two clusters are
$\lambda=87.8$ and $\lambda=84.0$ respectively, which demonstrates that
neither system is obviously dominant.  It is difficult to imagine a more spectacular
example of an SZ projection effect.


\subsubsection{Validation Tests for the \PSZ\ Redshifts}
\label{sec:PXXIXvalidation}

We use the $\lambda$--$M_{SZ}$ relation from Section \ref{sec:scaling} as a validation test of
the \PSZ\ cluster redshifts.  Specifically, let us assume that for each of our redshift conflicts,
the \PSZ\ redshifts are correct.
Papers I and II demonstrated that the failure rate of the \redmapper\ photometric redshifts
is below 1\%.  Consequently, if the \PSZ\ redshifts are correct, we ought to be able to
find good \redmapper\ matches to the \PSZ\ systems by redoing our rank-ordered circular
matching while adding the additional constraint that the \redmapper\ matches must be consistent ($3\sigma$)
with the \PSZ\ cluster redshifts.  In addition, based on our results in Section~\ref{sec:centering}, we only look for
cluster matches within $6\arcmin$ of the \Planck\ sources.  

Our visual inspection indicates that for the majority of the 35 redshift outliers identified in Section \ref{sec:matching},
the \PSZ\ redshifts are incorrect.  To test this, we perform the above matching --- including the redshift consistency
requirement --- and compare how these 35 galaxy clusters populate the $\lambda$--$M_{SZ}$ plane relative to 
the 185 galaxy clusters that defined $\lambda$--$M_{SZ}$ relation in Section~\ref{sec:scaling}.  Any \PSZ\ clusters
that go unmatched are arbitrarily assigned a richness $\lambda=5$, the minimum richness of the galaxy clusters
in our full \redmapper\ catalog.  


\begin{figure}
\hspace{-5mm} \includegraphics[width=90mm]{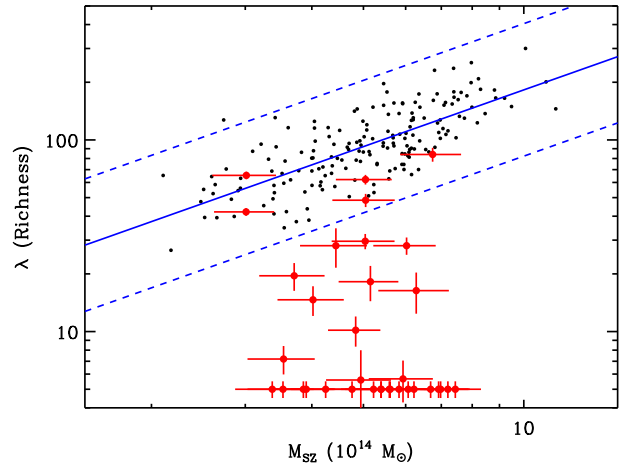}
\caption{Comparison of the location in the $\lambda$--$M_{SZ}$ plane for the 35 $3\sigma$ redshift outliers 
in Figure~\ref{fig:zcomp} (red points) and the points from Figure~\ref{fig:scaling} defining
the $\lambda$--$M_{SZ}$ relation.  The solid blue line is the mean scaling relation, and the dashed
lines delineate the $3\sigma$ scatter band.  The points at richness $\lambda=5$ are unmatched clusters.
To assign a richness to the \PSZ\ clusters, we have re-matched the redshift outliers
from Figure~\ref{fig:zcomp} while demanding that the \redmapper\ and \PSZ\ redshifts agree.  
In so doing, 30 clusters remain as outliers in the $\lambda$--$M_{SZ}$ plane, two of which
are due to bad SDSS photometry.  The rest are \PSZ\ redshift failures (see text).
} 
\label{fig:zout1}
\end{figure}


Our results are shown in Figure~\ref{fig:zout1}.  The solid blue line is the mean $\lambda$--$M_{SZ}$ relation (see
Fig.~\ref{fig:scaling}), while the dashed lines mark the $3\sigma$ band.  Of our 35 original outliers, only 5 clusters can plausibly
be matched assuming the \PSZ\ redshift.  In decreasing order of S/N, these are \Planck\ 732, 1128, 73, 510, and 216.
Turning to the results from our visual inspection (Section~\ref{sec:visual} and Table~\ref{tab:zout}), we see
that the first 4 systems were identified as SZ-projections of multiple clusters, while the last system is one for
which our original matching was incorrect.  
The IDs of the projection clusters are given in Table~\ref{tab:unmatched}.
The remaining 30 outliers break into 28 incorrect redshifts in \PSZ\ and two \redmapper\
failures (1 incompleteness, 1 bad photometry, \Planck\ 1216 and 500 respectively). 
 
We note that despite the fact that four of the five clusters we flagged as SZ projection are non-outliers in the $\lambda$--$M_{SZ}$ plane,
we should not consider this as evidence that the \PSZ \ redshift is appropriate. 
Indeed, in the following section, we find
that these systems do not appear as outliers when we remeasure $M_{SZ}$ using the assigned
\redmapper\ redshift either.    In other words, the fundamental problem for these systems is that these 
projection effects cannot be assigned a single, unambiguous redshift: 
the SZ detection is fundamentally a combination of more than one structure along the line-of-sight.


\subsubsection{Validation Tests for the \redmapper\ Redshifts}
\label{sec:RMvalidation}

We now perform the converse analysis to that of Section \ref{sec:PXXIXvalidation}, that is, we take
the 35 redshift conflicts from Figure~\ref{fig:zcomp} and assume that the automatically assigned \redmapper\
match is correct.  We then remeasure $M_{SZ}$ for each of these 35 systems, holding the \Planck\ location
fixed, but adopting the redshift of the assigned \redmapper\ cluster as the correct redshift.

Our visual inspection of the redshift outliers suggests that we will be able to confirm 19 cluster redshifts
as correct.  In addition, it is possible for the five systems labeled as SZ-projections to appear to be
acceptable cluster matches, for a total of 24 expected non-outliers.  
The remaining 11 systems (8 unconvincing clusters, 3 clusters where the \PSZ\ redshift is correct)
we expect will be outliers in the $\lambda$--$M_{SZ}$ plane.


\begin{figure}
\hspace{-5mm} \includegraphics[width=90mm]{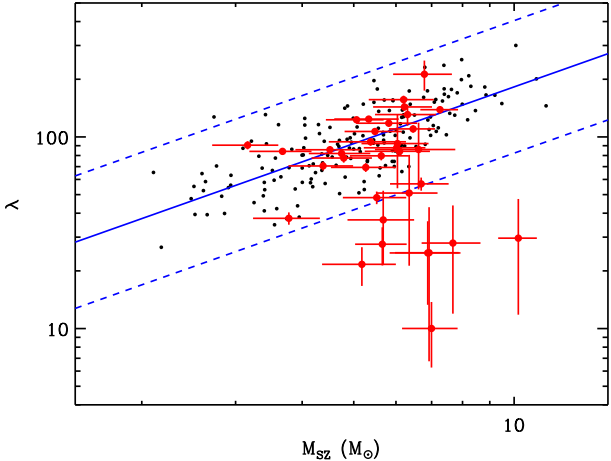}
\caption{Comparison of the location in the $\lambda$--$M_{SZ}$ plane for the redshift outliers 
in Figure~\ref{fig:zcomp} (red points) and the points from Figure~\ref{fig:scaling} defining
the $\lambda$--$M_{SZ}$ relation.  The clusters are assigned the richness of the best \redmapper\
match, and we have remeasured $M_{SZ}$ at the assigned \redmapper\ redshift for each of the
clusters.  Note a significant fraction of the clusters that were outliers in Figure~\ref{fig:zout1}
are not outliers in this figure.  This analysis confirms
the results of our visual inspection with regards to the incidence
of redshift failures in the \PSZ\ and the existence of SZ projections (see text for details). 
} 
\label{fig:zout2}
\end{figure}


As shown in Figure~\ref{fig:zout2}, this is almost exactly what we find.  
The solid blue line in the Figure is the mean $\lambda$--$M_{SZ}$ relation,
while the dashed lines mark the $3\sigma$ band.  Of the original 35 redshift outliers, nine fall below the
$3\sigma$ intrinsic scatter line.  These are the three clusters with correct \PSZ\ redshifts,
and six of the eight unconvincing cluster matches.   However, two of the
unconvincing matches are not obvious outliers in the $M_{SZ}$--$\lambda$ plane, and therefore
remain as plausible cluster matches.  These are cluster \Planck\ 376 and 768.    Note that \Planck\ 376 is
an unconvincing match in part because it may be an SZ projection.    

Turning to the remaining clusters labelled as projection effects, we see that these systems do not show up as outliers
in the $M_{SZ}$--$\lambda$ plane, so the scaling relation test is not able to rule out either the \PSZ\ or the
\redmapper\ redshifts.  Thus, the redshift association for these systems is ambiguous.
Importantly, the fact that these clusters are not outliers is {\it not} due to covariance of the type studied, e.g. in 
\citet{cohnwhite09}, \citet{nohcohn12}, \citet{anguloetal12}, or \citet{rozoetal12c}.  
These clusters are clear projections in the SZ, but they are not optical blends; the clusters are very well separated along the line-of-sight in the optical.

Table \ref{tab:newsz} below includes the 21 \PSZ\ clusters with revised redshifts that result in $\lambda$ and 
$M_{SZ}$ values consistent with the cluster scaling relations.  These are the 35-9=26 objects, minus the latter 
five labeled as projection effects.  In conjunction with the 185 galaxy clusters that originally defined the $\lambda$--$M_{SZ}$
scaling relation, these redshift-corrected systems brings the total number of galaxy clusters establishing the $\lambda$--$M_{SZ}$
relation to over 200 clusters.


\subsection{Summary of Results}
\label{sec:consolidation}

There are 245 \Planck\ SZ-detections in \PSZ\ labeled as confirmed that fall within the SDSS \redmapper\ footprint
with an assigned redshift $z$ in the range $z\in[0.08,0.6]$.  
These have allowed us to identify
three failures in the \redmapper\ cluster catalog.  One (\Planck\ 1216) is a low ($z\leq 0.6$) 
redshift cluster missing from the \redmapper\ catalog because of a combination of miscentering and angular masking,
reflecting $\approx 0.5\%$ incompleteness in \redmapper.  One (\Planck\ 500) is 
a redshift failure because of bad photometry in the SDSS, and one (\Planck\ 617, Section~\ref{sec:scaling}) has an unusually low richness because
of bad SDSS photometry.  Our estimate of the catastrophic failure rate for the SDSS \redmapper\ catalog from this analysis
is therefore $3/245\approx 1.2\%$.
In addition, we identified 4 \redmapper\ clusters as having two obvious galaxy concentrations, which can lead to
large centering uncertainties (\Planck\ 728, 249, 472, and 587, Section~\ref{sec:centering}), and one obvious 
centering failure in \redmapper\ (\Planck\ 113, Section~\ref{sec:centering}).  Note that the centering failure rate
of \redmapper\ is, in fact, higher, as evidenced by utilizing high resolution X-ray data \citep{rozorykoff13}.  The low
rate of miscentering identifications from this analysis reflects the fact that \Planck\ cluster centering is noisy.

Our analysis has also allowed us to flag 36 clusters that were assigned an incorrect redshift in \PSZ: 
five in section \ref{sec:scaling}, three in section \ref{sec:centering}, and 28 in sections \ref{sec:PXXIXvalidation}. 
For each of the clusters with incorrectly assigned redshifts in the \PSZ, we have assigned the correct cluster redshift where
possible, and remeasured $M_{SZ}$ at the newly assigned redshift. 
This is larger than the number of outliers in Figure \ref{fig:zcomp} because some of the clusters that do not
show up as outliers in that Figure have incorrect redshifts, as found in Sections~\ref{sec:scaling} and \ref{sec:centering}. 
The corresponding redshift failure rate in the \PSZ\ over 
the SDSS region is $36/245=14.7\%$.  These results are collected in Table~\ref{tab:newsz}.    

We caution that the \PSZ\ failure rate is not easily extrapolated outside the SDSS region, since some of the \PSZ\ failures
occur because of the reliance on existing optical catalogs in the SDSS region.  For instance, the 5 failures from section \ref{sec:scaling}
were matched to existing SDSS catalogs, and would have remained as cluster candidates (rather than confirmed clusters) 
had they fallen outside the SDSS
footprint.  We have investigated each of the failures identified in this work, and have found that roughly
a third of them tracked back to errors in either NED, SIMBAD, or the REFLEX cluster catalog.
Another third of the failures are systems in which the association made in the construction of the \PSZ\ cluster catalog was incorrect.
The remaining third are systems where the origin of the failure is unusual and/or not easily understood.  
We collect notes on each of the failures we identified in Appendix \ref{app:comments}.


\begin{table}
\centering
\caption{Revised cluster redshifts and SZ masses for PXXIX Clusters with incorrect cluster redshifts.
Clusters tagged with an asterisk$^*$ are systems that are not outliers in the $\lambda$--$M_{SZ}$ plane
given our assigned redshift, but which were labelled as unconvincing cluster matches based on our visual
inspection.}
\begin{tabular}{rccr}
\hline
Cluster & Redshift & $M_{SZ}\ (10^{14}\ M_{\sun})$ & $\lambda \hspace{0.25in}$ \\
\hline
513\hspace{0.0in} & 0.363 $\pm$ 0.014 & 7.26 $\pm$ 0.59 & 138.8 $\pm$ 6.1 \\
391\hspace{0.0in} & 0.276 $\pm$ 0.009 & 6.46 $\pm$ 0.64 & 110.1 $\pm$ 4.8 \\
660\hspace{0.0in} & 0.297 $\pm$ 0.011 & 5.48 $\pm$ 0.67 & 106.8 $\pm$ 4.4 \\
316\hspace{0.0in} & 0.354 $\pm$ 0.014 & 6.22 $\pm$ 0.79 & 143.4 $\pm$ 6.4 \\
537\hspace{0.0in} & 0.289 $\pm$ 0.010 & 5.06 $\pm$ 0.63 & 122.9 $\pm$ 4.7 \\
888\hspace{0.0in} & 0.327 $\pm$ 0.013 & 5.63 $\pm$ 0.71 & 79.7 $\pm$ 4.2 \\
222\hspace{0.0in} & 0.163 $\pm$ 0.004 & 3.67 $\pm$ 0.49 & 84.1 $\pm$ 3.4 \\
865\hspace{0.0in} & 0.233 $\pm$ 0.007 & 4.74 $\pm$ 0.63 & 82.0 $\pm$ 4.0 \\
303\hspace{0.0in} & 0.264 $\pm$ 0.008 & 4.79 $\pm$ 0.62 & 77.6 $\pm$ 4.2 \\
97\hspace{0.0in} & 0.307 $\pm$ 0.013 & 5.52 $\pm$ 0.75 & 48.2 $\pm$ 3.6 \\
668\hspace{0.0in} & 0.334 $\pm$ 0.014 & 5.82 $\pm$ 0.76 & 117.9 $\pm$ 5.0 \\
443\hspace{0.0in} & 0.221 $\pm$ 0.006 & 3.16 $\pm$ 0.45 & 90.6 $\pm$ 4.4 \\
764\hspace{0.0in} & 0.466 $\pm$ 0.011 & 6.31 $\pm$ 0.83 & 130.8 $\pm$ 16.5 \\
1123\hspace{0.0in} & 0.219 $\pm$ 0.006 & 5.27 $\pm$ 0.74 & 69.3 $\pm$ 3.7 \\
308\hspace{0.0in} & 0.197 $\pm$ 0.005 & 4.51 $\pm$ 0.62 & 85.7 $\pm$ 3.9 \\
$768^*$\hspace{0.0in} & 0.326 $\pm$ 0.013 & 5.38 $\pm$ 0.89 & 94.3 $\pm$ 4.3 \\   
779\hspace{0.0in} & 0.296 $\pm$ 0.010 & 5.33 $\pm$ 0.74 & 123.9 $\pm$ 4.8 \\
292\hspace{0.0in} & 0.229 $\pm$ 0.007 & 4.38 $\pm$ 0.61 & 70.6 $\pm$ 3.7 \\
$376^*$\hspace{0.0in} & 0.161 $\pm$ 0.005 & 3.78 $\pm$ 0.54 & 37.6 $\pm$ 2.8 \\   
416\hspace{0.0in} & 0.372 $\pm$ 0.014 & 6.20 $\pm$ 0.87 & 156.5 $\pm$ 7.6 \\
13\hspace{0.0in} & 0.327 $\pm$ 0.013 & 6.10 $\pm$ 0.85 & 84.1 $\pm$ 4.3 \\
\hline
\end{tabular}
\label{tab:newsz}
\end{table}


It is important to note, however, that some clusters remain with uncertain redshifts, as we were neither able
to confirm the \PSZ\ redshift as correct, nor to assign a new redshift to the \Planck\ SZ source.
In some cases, this is because the \Planck\ SZ detection is associated with two
comparably rich clusters that are projected along the line of sight, making a unique association
impossible.  In other cases, the \Planck\ sources have no acceptable \redmapper\
counterpart.  In the latter case, we have classified these systems
as either high redshift cluster candidates or false detections based on visual inspection of SDSS and WISE data.
The set of clusters for which we do not assign a robust redshift is
collected in Table~\ref{tab:unmatched}. 


\begin{table*}
\centering
\caption{PXXIX clusters with no unique good redshift match.  For high-$z$ cluster candidates, the RA and DEC columns
contain the position of the optical cluster candidate optical, and its redshift where available.  The redshift quoted is always
the SDSS DR8 photometric redshift of the central galaxy, except for cluster \Planck\ 299, where the redshift is spectroscopic.}
\begin{tabular}{rcccrrc}
\hline
Cluster & SZ Projection & Non-Detection &Confidence & RA & DEC & $z$ \\
	&	& High-$z$ Candidate & & & \\
\hline

299\hspace{0.0in} &  ---  &  ---   & 5   & 231.6383 & 54.1520 & $\zspec=0.748$ \\			
795\hspace{0.0in} &  ---  &  ---   & 5  & 170.4636 & 15.8030 & $0.77 \pm 0.07$ \\		
441\hspace{0.0in} &  ---  &  ---   & 5  & 10.7735 & 18.3686 & $0.63 \pm 0.04$ \\		
400\hspace{0.0in} &  ---  &  ---   & 4  & 3.0807 & 16.4621 & $0.64\pm 0.08$ \\			
574\hspace{0.0in} &  ---  &  ---   & 3   & 170.9893 & 43.0600 & $0.72\pm 0.13$ \\
280\hspace{0.0in} &  ---  &  ---   & 3   & 340.6030 & 17.5214 & $0.86\pm 0.10$ \\
650\hspace{0.0in} &  ---  &  ---   & 3  & 176.5397 & 30.8911 & $0.60\pm 0.17$ \\		
664\hspace{0.0in} &  ---  &  ---   & 3  & 143.4653 &  28.0855 & $0.60 \pm 0.07$  \\		
505\hspace{0.0in} &  ---  &  ---   & 2   & 172.23425 & 59.94395 & $0.64 \pm 0.08$ \\
527\hspace{0.0in} &  ---  &  ---   & 2   & 21.71503 & -7.20770 & $0.72 \pm 0.04$ \\  
719\hspace{0.0in} &  ---  &  ---   & 2   & 158.6913 & 20.5628 & --- \\	
768\hspace{0.0in} &  ---  &  ---   & 2   & 158.28934, & 13.79361 & $0.58 \pm 0.13$ \\					
622\hspace{0.0in} &  ---  & ---  & 1  & 42.8736 & -7.9581 & --- \\
484\hspace{0.0in} &  ---   & --- & 1  & 178.1366 & 61.3629  & $0.86\pm 0.11$ \\
678\hspace{0.0in} &  ---  &  ---   & 1  & 147.0918 & 24.7901 & $ 0.59 \pm 0.04$ \\
1093\hspace{0.0in} &  ---  &  ---   & 1 & 193.8617 & 21.1175 & --- \\ 					
729\hspace{0.0in} &  ---  &  ---   & 1 & 140.6389 & 11.6842 & --- \\					
732\hspace{0.0in} &  $\checkmark$  &  ---   & ---   & --- & --- & --- \\
1128\hspace{0.0in} &  $\checkmark$  &  ---   & ---   & --- & --- & --- \\
73\hspace{0.0in} &  $\checkmark$  &  ---   & ---   & --- & --- & --- \\
510\hspace{0.0in} &  $\checkmark$  &  ---   & ---   & --- & --- & --- \\
234\hspace{0.0in} & $\checkmark$ & --- & --- & --- & --- & --- \\
376\hspace{0.0in} &  ?  &  ?   & ---   & --- & --- & --- \\
77\hspace{0.0in} &  ---  &  ?   & ---   & --- & --- & --- \\
52\hspace{0.0in} &  ---  &  $\checkmark$   & ---   & --- & --- & --- \\
\hline
\end{tabular}
\label{tab:unmatched}
\end{table*}



\section{Discussion}
\label{sec:conclusions}

We have performed a detailed comparison of the SDSS DR8 \redmapper\ and \PSZ\ cluster catalogs
that has enabled us to characterize systematic failures in both.  The resulting
failure rates are $1.2\%$ for \redmapper\ and $14.7\%$ for the \PSZ\ over the SDSS area.  A summary of the various
failure modes is presented in Section~\ref{sec:consolidation}.  

These results firmly demonstrated that robust photometric cluster finding is possible, 
and that state-of-the-art photometric cluster finding algorithms like \redmapper\
can deliver the necessary cluster samples to exploit near future photometric surveys.
Moreover, we have confirmed
optical cluster richness as a relatively low-scatter mass proxy $(\sigma_{\ln M|\lambda}\approx 0.21$), a result
that has now been independently established with three different mass proxies: $M_{gas}$, $T_X$, and $M_{SZ}$.
We note that
this scatter in mass at fixed observable for cluster richness is comparable to that which can be
attained using X-ray or SZ survey-quality data.  Of course, pointed follow-up X-ray observations with high resolution
instruments such as {\it Chandra} and {\it XMM} remains valuable, and the corresponding mass proxies exhibit
lower scatter. 

Our findings do not affect the \Planck\ cluster cosmology analysis presented in \citet{pXX}. 
We found only one object over the SDSS area from the cosmology cluster sample with an 
incorrect redshift; the vast majority of the redshift failures in the \PSZ\ are lower S/N systems that have 
not yet been included as part of the cosmological analysis.
In addition, it is not clear
to what extent the \PSZ\ failure rate can be extended outside the SDSS region.  The Northern extragalactic sky, 
and in particular the SDSS area, has been extensively studied, and many of the
\PSZ\ redshift failures originate in identifications with catalogs not available elsewhere on the sky.  
In short, the failure rate in the \PSZ\ 
calculated in this work is at least in part
due to failures in existing data bases and/or optical cluster catalogs, or failures due to the difficulty of performing unambiguous
associations between the \Planck\ SZ detections and other optical catalogs.

Indeed, this is one of more significant aspects of our results. PXXIX found it difficult to
capitalize on available optical/IR cluster samples for validation of the \PSZ\ because of 
the large scatter in the richness--$M_{SZ}$ relation.
Our results unambiguously
demonstrate that the difficulties encountered by PXXIX are not due to a generic
feature of photometric cluster catalogs;
one can construct high-quality photometric catalogs for which scaling relations may be used
to unambiguously pair optical and SZ cluster samples.
Our success with \redmapper\ is of particular importance given that we are about to usher in a new era of
large photometric surveys, which include the Dark Energy Survey \citep[DES,][]{des05},
the Hyper-Suprime Camera (HSC) survey, the Large Synoptic Survey Telescope \citep[LSST,][]{lsst09},
the \Euclid\ mission and WFIRST.   

Our results also highlight the importance of multi-wavelength data for cluster cosmology.  As in PXXIX and Paper II,
we have found that comparing clusters catalogs selected in different wavelengths can help identify previously
unaccounted for systematics.  Just as importantly, we find that
optical survey data can inform X-ray/SZ cluster catalogs just as much as X-ray/SZ data can help inform
optical cluster catalogs.  Thus, our success with \redmapper\ is important not only from 
the point of view of utilizing future photometric surveys for cluster cosmology,
it also highlights the important role that these surveys will have on future SZ and X-ray clusters samples.
In short, our results clearly demonstrate that
there is good reason to believe that surveys like DES, HSC, LSST, \Euclid\ and WFIRST  will play 
an important role in the development of cluster science {\bf together} with X-ray and SZ surveys such as {\it eRosita}, \Planck,
the Atacama Cosmology Telescope (ACT), and the South Pole Telescope (SPT).

Our results also highlight the importance of multi- wavelength data for cluster cosmology. As in PXXIX and Paper II, we have 
found that comparing clusters catalogs selected in different wavelengths can help identify previously unaccounted for systematics.  
The improved understanding of cluster selection and characterization gained by  combining redMaPPer and Planck analyses 
demonstrates the intrinsic value to cluster science in combining next generation optical-IR 
(DES, HSC, LSST, Euclid and WFIRST) surveys with X-ray (XCS, XMM-XXL, eRosita) and SZ 
(Planck, Atacama Cosmology Telescope, South Pole Telescope) surveys.

\section*{Acknowledgments} 

We thank August Evrard for comments on an early draft of this manuscript.
This work was supported in part by  the U.S. Department of Energy contract to SLAC no. DE-AC02-76SF00515.
JGB gratefully acknowledges support from the {\it Institut Universitaire de France}. A portion of the research described 
in this paper was carried out at the Jet Propulsion Laboratory, California Institute of Technology, under a 
contract with the National Aeronautics and Space Administration. 


\newcommand\AAA{{A\& A}}
\newcommand\PhysRep{{Physics Reports}}
\newcommand\apj{{ApJ}}
\newcommand\PhysRevD[3]{ {Phys. Rev. D}} 
\newcommand\prd[3]{ {Phys. Rev. D}} 
\newcommand\PhysRevLet[3]{ {Phys. Rev. Letters} }
\newcommand\mnras{{MNRAS}}
\newcommand\PhysLet{{Physics Letters}}
\newcommand\AJ{{AJ}}
\newcommand\aap{ {A \& A}}
\newcommand\apjl{{ApJ Letters}}
\newcommand\aph{astro-ph/}
\newcommand\AREVAA{{Ann. Rev. A.\& A.}}
\newcommand\pasj{PASJ}
\newcommand\apjs{{ApJ Supplement}}
\newcommand\jcap{JCAP}

\bibliographystyle{mn2e}
\bibliography{mybib,planck}

\appendix


\section{Comments on Individual Clusters}
\label{app:comments}

{\it Planck 13:}  This is a redshift outlier.  The SZ source is associated with ZwCl 1454.5+0656 at 2\arcmin\ from the \Planck\ position.  The redshift is associated with NSC J145659+064610 at 2.3\arcmin, and these two clusters appear to be one and the same.  Thus, the association
in the \PSZ\ is correct, but the NED photometric redshift is incorrect, as can be confirmed with SDSS spectroscopy
($\zspec=0.312$).
There is a \citet{wenetal12} object close to the \redmapper\ object, at 2.3\arcmin, with the same redshift.  
\vspace{0.05in}

{\it Planck 73: } The \Planck\ location is roughly half way between two modestly rich \redmapper\ clusters,
RM 1130 ($\lambda=42.1$, $\zspec=0.091$, $\Delta\theta=3.4\arcmin$) and
RM 4408 ($\lambda=56.9$, $\zspec=0.386$, $\Delta\theta=2.0\arcmin$).  As such, it appears
to be a clear case of an SZ projection.  The cluster is not an outlier in the $M_{SZ}$--$\lambda$
plane when adopting either redshift, demonstrating that there is no unique redshift that can be associated with this
detection.
\vspace{0.05in}

{\it Planck 77: } PXXIX associated this cluster with RXC J1453.1+2153 and A1986 in the \PSZ, at z=0.1186 and 7.8\arcmin\  from the \Planck\ position (Section~\ref{sec:centering}).  There is no good \redmapper\ counterpart.  The \Planck\ detection is made by a single algorithm at a $S/N=4.58$.  The positional shift between \Planck\ and A1986 corresponds to a separation of $\sim \theta_{500}$, a significant shift, but one that has been seen before for low-redshift clusters with large images on the sky.  We tentatively consider this a false detection, though this classification is insecure.  
\vspace{0.05in}

{\it Planck 97: } This is a redshift outlier.  The \Planck\ location is roughly half way between two rich \redmapper\ clusters,
RM 19525 ($\lambda=28.0$, $z_\lambda=0.357\pm 0.016$, $\Delta\theta=3.5\arcmin$) and
RM 4470 ($\lambda=48.2$, $\zspec=0.310$, $\Delta\theta=4.3\arcmin$).  PXXIX assigned
this SZ source the redshift of the closer, less massive clusters, which results in an outlier in the 
$\lambda$--$M_{SZ}$ plane.  The source is {\it not} an outlier when adopting RM 4470 as the correct
cluster match, so we adopt this match as correct. 
\vspace{0.05in}

{\it Planck 216: } Our automated matching algorithm matched this \Planck\ cluster to RM 96829
($\lambda=50.9$, $z_\lambda=0.600$, $\Delta\theta=6.0\arcmin$).  
A second, obviously correct candidate is RM 5279
($\lambda=48.5$, $\zspec=0.336$, $\Delta\theta=1.5\arcmin$).  This latter choice corresponds
to the redshift assigned by PXXIX, so this
cluster is one of the redshift outliers introduced by a failure of the matching algorithm.
\vspace{0.05in}

{\it Planck 222:}  This is a redshift outlier. The SZ source is associated with RXC J1421.6+3717 in the \PSZ\ and the redshift taken from the MCXC.  The position also matches A1902 and an object from the \citet{wenetal12} catalog.  The latter two objects have redshifts compatible with the \redmapper\ value of $z=0.162\pm 0.005$.  The \Planck\ association is correct, but the redshift from the MCXC is incorrect.
\vspace{0.05in}

{\it Planck 234: } The \Planck\ source sits almost directly in between two \redmapper\
clusters, RM 24517 
($\lambda=92.4$, $z_\lambda=0.584$, $\Delta\theta=2.5\arcmin$) and RM 24933
($\lambda=28.6$, $z_\lambda=0.594$, $\Delta\theta=3.3\arcmin$).  Note the richness values
are highly uncertain because of the high redshift.  Given that the SZ detection is
nearly in the middle of the two systems, and that the redshift values are unreliable,
we consider this an SZ-projection effect.
It is notable that a third system with $\lambda=20.0$ at $z_\lambda=0.470$
and $\Delta\theta=2.0\arcmin$ is also present in the field.
\vspace{0.05in}

{\it Planck 280: }  This cluster was associated with RXC J2241.8+1732 and A2472 in the \PSZ, at z=0.3137 and 7.2\arcmin\ from the \Planck\ position (Section~\ref{sec:centering}).  This is also the closest candidate \redmapper\ match.  
This corresponds to a rather large shift of $\sim 2\theta_{500}$.  There is a 
possible match to an over-density in SDSS and WISE at $RA=340.6030$ and $DEC=17.5214$ 
(2.2\arcmin\ separation) and $z=0.86$.  May be a projection of two SZ sources.
\vspace{0.05in}

{\it Planck 292:}  This is a redshift outlier. The SZ source is associated with ZwCl 1341.2+4022 in the \PSZ\ at 2.9\arcmin\ from the \Planck\ position, which has no redshift in NED.  The redshift given in the \PSZ\ is $z=0.186$ is that of GMBCG J205.85324+40.09379 at 1.5\arcmin.
\vspace{0.05in}

{\it Planck 299:}  Outlier in the $\lambda-M_{SZ}$ relation (Section~\ref{sec:scaling}).  The SZ source is associated with AMF J231.538+54.1303 in the \PSZ, at z=0.4735 and 2.9\arcmin\ from the \Planck\  position.  It is also identified with the low-richness clusters WHL ($N_{200}=15$) and GMBCG $N_{\rm scaled-gals}=13$.  It has a \redmapper\ counterpart at the same redshift, but of too low richness to be an acceptable match.  There is an SDSS over-density at 3\arcmin\ from the \Planck\ position, at $RA=231.6383$ and $DEC=54.1520$, with a spectroscopic redshift $\zspec=0.748$ (Fig.~\ref{fig:299}, and Table \ref{tab:unmatched}).  This is a candidate high-z cluster that appears to have been incorrectly associated with a low-mass system in the foreground.
\vspace{0.05in}

{\it Planck 303:}  This is a redshift outlier. The SZ source is associated with ZwCl 2341.1+0000 in the \PSZ\ at 2.5\arcmin\ from the \Planck\ position.  This object is also known as SDSS CE J355.930756+00.303606 and NSCS J234339+001747 with redshift compatible with the \redmapper\ value of $z=0.274\pm009$.  The redshift given in the \PSZ\ comes from SIMBAD.  The \Planck\ association appears correct, but the SIMBAD redshift inaccurate.
\vspace{0.05in}

{\it Planck 308:}  This is a redshift outlier. The SZ source is associated with ACO 2623 in the \PSZ\ at 2.2\arcmin\ from the \Planck\ position.  The ACO cluster lies between the \Planck\ SZ source and the \redmapper\ cluster at 2.6\arcmin\ from the \Planck\ position.  The redshift given in NED for ACO 2623 is based on a single galaxy redshift.
\vspace{0.05in}

{\it Planck 316:}  This is a redshift outlier.  The SZ source is associated with ZwCl 2341.8+0251 at 4.1\arcmin\  separation, but the redshift reported in the \PSZ\ ($z=0.43$) corresponds to NSCS J234433+030506 at a separation of 1.5\arcmin\ from the \Planck\ position.  The \redmapper\ counterpart lies at 1.2\arcmin\ from the \Planck\ position and also from the NSCS object, which is also clearly the same system as the \redmapper\ object.
The redshift from NED is photometric, and incorrect, as confirmed with SDSS spectroscopy ($\zspec=0.350$).
\vspace{0.05in}

{\it Planck 376: } The \Planck\ source sits almost directly in between two \redmapper\
clusters, RM 4427
($\lambda=37.6$, $\zspec=0.159$, $\Delta\theta=5.8\arcmin$) and RM 27136
($\lambda=14.6$, $\zspec=0.178$, $\Delta\theta=5.3\arcmin$).  The second system
is the one that PXXIX associated with \Planck\ 376, and is an outlier in the $\lambda$--$M_{SZ}$
relation.  Given the modest richness
of the two clusters, and the relatively large angular offsets, neither cluster match
is particularly convincing.  This appears to be either a false detection, or an SZ-projection
effect.
\vspace{0.05in}

{\it Planck 391:}  This is a redshift outlier.  The SZ source is associated with ZwCl 0017.0+0320 in the \PSZ, at 0.2\arcmin\ separation, which
also corresponds to the correct \redmapper\ match.  The quoted photometric redshift in NED utilized by PXXIX is incorrect.
\vspace{0.05in}

\noindent  {\it Planck 400:} This cluster has no acceptable \redmapper\ match. 
Visual inspection of SDSS and WISE reveals a rich high redshift cluster candidate at
$RA=3.08068$, $DEC=16.46209$, $z=0.64\pm 0.08$.  
\vspace{0.05in}

{\it Planck 416:}  This is a redshift outlier.  The SZ source is associated with RXC J0019.6+2517 with $z=0.1353$ taken from the NED database.  The RXC and \redmapper\ counterpart, at $z=0.373\pm 0.014$, are separated by only 0.8\arcmin.  There is another \redmapper\ system close to the RXC with the redshift given by NED, but it has a low richness, $\lambda=7.1$, which is to be compared with $\lambda=156.5$ for the higher redshift system.  It is possible that the association between the X-ray source and the Planck SZ detection is correct, and that the X-ray source was incorrectly associated with the foreground group rather than the background cluster that we identify as the proper counterpart to the \Planck\  SZ source.
\vspace{0.05in}

{\it Planck 441:}   Outlier in the $\lambda-M_{SZ}$ relation (Section~\ref{sec:scaling}).  The SZ source is associated with the low-richness cluster WHL J10.6526+18.4259 ($N_{200}=14$) in the \PSZ, at $z=0.2668$ and 6.5\arcmin\ from the \Planck\  position.  It has a \redmapper\ counterpart at the same redshift, but of too low richness.  There is a WISE over-density with faint optical counterpart located 1.4\arcmin\ from the \Planck\ position, at $RA=10.7735$ and $DEC=18.3686$, with estimated $z=0.63$ (Table \ref{tab:unmatched}).  This is a candidate high-z cluster that has been incorrectly associated with a low-mass system in the foreground.
\vspace{0.05in}

{\it Planck 443: } Our automated matching algorithm matched this cluster to RM 319 
($\lambda=90.6$, $\zspec=0.221$, $\Delta\theta=5.2\arcmin$).  
A second candidate is RM 49592
($\lambda=28.1$, $z_\lambda=0.438$, $\Delta\theta=0.4\arcmin$).  This is the association
made by PXXIX.  Neither candidate is obviously correct based on our visual inspection.
The cluster is an outlier in $\lambda$--$M_{SZ}$ when paired with RM 49592, but is not an outlier
when paired with RM 319.   We make this association, noting the cluster likely suffers from SZ projection.
\vspace{0.05in}

{\it Planck 484:}  This is a redshift outlier.  The SZ source is associated with WHL J178.058+61.33, with $z=0.3169$ at 1.8\arcmin.  This is a low-richness system ($N_{200}=11$) that is unlikely to be the \Planck\ counterpart.  There is no good \redmapper\ counterpart, but we were able to 
identify a possible high-z
match at $RA=178.1366$ and $DEC=61.3629$ with $z=0.86\pm0.11$ (see Table~\ref{tab:zout}).
\vspace{0.05in}

{\it Planck 505:} This is a redshift outlier.  The SZ source is detected by only one (MMF3) of the three \Planck\ methods, at $S/N=4.53$.  The redshift comes from dedicated follow-up observations.  The cluster has no good \redmapper\ counterpart, but we were able to identify a candidate
high redshift galaxy overdensity.
\vspace{0.05in}

{\it Planck 510: } This cluster sits directly on top of 2 rich ($\lambda=87.8$ and $84.0$) galaxy clusters 
at redshifts $\zspec=0.256$ and $\zspec=0.373$ respectively.  The corresponding angular separations
relative to the \Planck\ location are $2.0\arcmin$ and $0.6\arcmin$.  
This is a spectacular example of an SZ-projection effect.
\vspace{0.05in}

{\it Planck 513:} This is a redshfit outlier.  The SZ source is associated with RXC J1159.2+4947, also A1430, in the \PSZ.  The redshift for the counterpart was taken from NED and SIMBAD, both of which incorrectly give $z=0.211$.  The \PSZ\ association is correct, but the redshift incorrect.  This cluster is part of the \Planck\ cluster cosmology sample.
\vspace{0.05in}

{\it Planck 527:}  This is a redshift outlier.  The SZ source is detected by only one (MMF1) of the three \Planck\ detection methods, at $S/N=4.52$.  It's redshift comes from dedicated follow-up observations.
The cluster has no good \redmapper\ counterpart, but we were able to identify a candidate
high redshift galaxy overdensity.
\vspace{0.05in}

{\it Planck 537:}  This is a redshift outlier.  The SZ source is associated with RXC J1017.5+5934, also A0959.  The redshift assigned in the \PSZ\ ($z=0.353$) is given by SIMBAD.  The redshift for the same object in NED is $z=0.2883$, in agreement with the \redmapper\ match.  The \Planck\ association appears correct, but with an incorrect redshift from SIMBAD.
\vspace{0.05in}

{\it Planck 574: } The \Planck\ cluster was originally matched to RM 97769 
($\lambda=36.9$, $z_\lambda=0.530$, $\Delta\theta=7.1\arcmin$).  The cluster
is clearly properly centered in the optical, so the large angular separation suggests this
is not a good match.  A second candidate is RM 4112
($\lambda=19.5$, $z_\lambda=0.192$, $\Delta\theta=3.0\arcmin$).  This is the association
made by PXXIX, but we note this cluster
is less rich than all 185 good cluster matches shown in Figure~\ref{fig:scaling}, and that this
association results in an outlier in the $\lambda$--$M_{SZ}$ plane.
This suggests the cluster is not rich enough to lead to a \Planck\ detection.  SDSS and WISE images both
reveal a candidate high redshift system at $z\approx 0.7$ within $2.9\arcmin$, making this
a good high redshift cluster candidate.
\vspace{0.05in}

{\it Planck 622:}  This is a redshift outlier.  The redshift assigned by \Planck\ was obtained in dedicated follow-up imaging with the Russian Turkish Telescope.
The cluster has no good \redmapper\ counterpart, but we were able to identify a candidate
high redshift galaxy overdensity, albeit with low confidence.
\vspace{0.05in}

{\it Planck 660:}  This is a redshift outlier.  The SZ source is correctly associated with ZwCl 0928.0+2904, which NED reports has a photometric redshift $z=0.3185$ (SDSS gives a spectroscopic redshift $z_\mathrm{spec}=0.2959$), and is at 1.8\arcmin\ from the \Planck\ position.  However, the redshift assigned in the \PSZ\ is $z=0.222$, corresponding to the cluster GMBCG J142.68564+28.82848, which is incorrect.
\vspace{0.05in}

{\it Planck 668:}  This is a redshift outlier. The SZ source is associated with ZwCl 0824.5+2244 in the \PSZ\ at 0.66\arcmin\ from the \Planck\ position with $z=0.3175$, compatible with the \redmapper\ value of $z=0.329\pm 0.013$.  SDSS spectroscopy confirms $\zspec=0.335$.
The \PSZ\ redshift of $z=0.287$ corresponds to that of NSC J082722+223244 at 2.9\arcmin\ from the \Planck\ position.  This last incorrect
redshift comes rom NED, where it is reported as a photometric redshift.  
The \Planck\ association is correct, but the redshift is incorrectly assigned.
\vspace{0.05in}

\noindent {\it Planck 678: } 
There are no good matching candidates for this system in \redmapper.  
Visual inspection reveals a very bright WISE source at ($RA=147.0918,DEC=24.7901$), also detected in SDSS, 
with a photometric redshift $z=0.59\pm 0.04$.  Several other candidate cluster galaxies can be discerned in the WISE image, making this a high redshift cluster candidate.  The SZ source is only detected by one (MMF1) of the three \Planck\  methods.  
\vspace{0.05in}

{\it Planck 719:}  Outlier in the $\lambda-M_{SZ}$ relation (Section~\ref{sec:scaling}).  The SZ source is associated with the low-richness cluster WHL J158.665+20.5346 ($N_{200}=11$) in the \PSZ, at $z=0.4674$ and 1.2\arcmin\ from the \Planck\  position.  It has a \redmapper\ counterpart at the same redshift, but of too low richness to be an acceptable match.  There is a WISE over-density with faint optical counterpart located 1.5\arcmin\ from the \Planck\ position, at $RA=158.6913$ and $DEC=20.5628$, of unknown redshift (Table \ref{tab:unmatched}).  This is a candidate high-z cluster that appears to have been incorrectly associated with a low-mass system in the foreground.
\vspace{0.05in}

{\it Planck 729:}  This cluster is an outlier in the $\lambda-M_{SZ}$ relation (Section~\ref{sec:scaling}).  The SZ source was associated with the low-richness cluster  WHL J140.630+11.6581 ($N_{200}=21$) in the \PSZ, at $z=0.2609$ and 2.3\arcmin\ from the \Planck\  position; also listed in the maxBCG and GMBCG catalogs.  It has a \redmapper\ counterpart at the same redshift, but of too low richness. 
There is a WISE over-density with faint optical counterpart located 0\arcmin.8 from the \Planck\ position, at $RA=140.6389$ and $DEC=11.6842$, of unknown redshift (Table \ref{tab:unmatched}).  This is a candidate high-z cluster that has been incorrectly associated with a low-mass system in the foreground.
\vspace{0.05in}

{\it Planck 732: } The \Planck\ location is roughly half way between two rich \redmapper\ clusters,
RM 1057 ($\lambda=62.1$, $z_\lambda=0.234$, $\Delta\theta=3.3\arcmin$) and 
RM 373 ($\lambda=212.4$, $z_\lambda=0.477$, $\Delta\theta=3.0\arcmin$).  As such, it appears to
be a clear case of an SZ-projection.  If we were to choose a single match, the better match is RM 373,
which is both richer and closer to the \Planck\ detection, whereas PXXIX assigned this detection
the redshift of the smaller, slightly more distant cluster RM 1057.  That said, we consider this cluster
an SZ projection with no unambiguous redshift.   Neither of the two matchings noted above results
in the cluster being an outlier in the $\lambda$--$M_{SZ}$ plane.
\vspace{0.05in}

{\it Planck 764:}  This is a redshift outlier. The \PSZ\ associated this SZ source with ZwCl 0924.4+0511 at 2.7\arcmin\ from the \Planck\ position (with $z=0.27$), but the given redshift of $z=0.2845$ is that of NSC J092656+045928, at 1.6\arcmin.  
The two clusters should be identified.  There is also a low-richness \redmapper\ cluster
at this redshift ($\lambda=29.6$, $z_\lambda=0.263\pm0.009$), consistent with SDSS spectroscopy for the group ($z_\mathrm{spec}=0.264$).  However,
the correct \redmapper\ match is a higher redshift system ($\lambda=130.8$, $z_\lambda=0.466\pm 0.011$).  SDSS spectra confirms the cluster's
redshift as $z_\mathrm{spec}=0.462$.
\vspace{0.05in}

{\it Planck 768: } The \Planck\ cluster was originally matched to RM 399
($\lambda=94.3$, $\zspec=0.315$, $\Delta\theta=5.8\arcmin$).  There are no other
plausible \redmapper\ matches.   This association does not make the cluster an outlier in the
$\lambda$--$M_{SZ}$ plane, but nevertheless, the angular offset is uncommonly
large.  Inspection of the SDSS and WISE images reveal a possible high redshift 
cluster match.
\vspace{0.05in}

{\it Planck 779:}  This is a redshift outlier. The SZ source is associated with ACO 1127 in the \PSZ\ at 4\arcmin\ from the \Planck\ position (with $z_\mathrm{photo}=0.3294$), but the given redshift $z_\mathrm{photo}=0.2564$ of NSC J105417+143835 at 1.8\arcmin.
The two clusters are one and the same, and they are the correct cluster match to the SZ source,
but neither redshift is correct.  SDSS spectroscopy confirms $\zspec=0.299$,
in agreement with the \redmapper\ redshift $z_\lambda=0.296\pm 0.010$.
\vspace{0.05in}

\noindent {\it Planck 795:} This SZ source is not a good match to any \redmapper\ clusters.
Visual inspection reveals a rich high redshift cluster at 
$RA=170.46358$, $DEC=15.80301$, $z=0.77 \pm 0.07$.  The galaxy overdensity is confirmed with WISE.
\vspace{0.05in}

{\it Planck 865:}  This is a redshift outlier. The SZ source is associated with A1208 in the \PSZ.  There is no redshift given in SIMBAD.  The redshift given in NED is compatible with the \redmapper\ $z=0.234\pm0.007$, as is the redshift of a nearby \citet{wenetal12} object.  The \Planck\ association appears correct, but the redshift is incorrect.
\vspace{0.05in}

{\it Planck 888:}  This is a redshift outlier. The SZ source is associated with ZwCl 1232.1+2319 at 2.7\arcmin\ from the \Planck\  position, but the redshift assigned in the \PSZ\ is $z_\mathrm{photo}=0.4125$, corresponding to the cluster NSC J123444+230059 at 0.451\arcmin.  
The 2 clusters appear to be one and the same, and they correspond to the correct cluster match,
but the redshift is incorrect.  The \redmapper\ object lies at 0.67\arcmin, and appears to be properly
centered.  SDSS spectra gives $\zspec=0.323$, confirming the \redmapper\ redshift $z_\lambda=0.327\pm 0.013$.
\vspace{0.05in}

{\it Planck 1093:}  Outlier in the $\lambda-M_{SZ}$ relation (Section~\ref{sec:scaling}).  The SZ source is associated with the low-richness cluster GMBCG J193.82512+21.04201 ($N_{scaled-gals}=8$) in the \PSZ, at $z=0.4457$ and 4.8\arcmin\ from the \Planck\  position.  It has a \redmapper\ counterpart at the same redshift, but of too low richness.  There is a WISE over-density with faint optical counterpart located 0\arcmin.2 from the \Planck\ position, at $RA=193.8617$ and $DEC=21.1175$, of unknown redshift (Table \ref{tab:unmatched}).  This is a candidate high-z cluster that appears to have been incorrectly associated with a low-mass system in the foreground.
\vspace{0.05in}

{\it Planck 1123: }  Our automated algorithm matched this cluster to RM 897
($\lambda=69.3$, $\zspec=0.233$, $\Delta\theta=5.1\arcmin$).    This tentative association
is not an outlier in the $\lambda$--$M_{SZ}$ plane.  
PXXIX assigned a redshift $z=0.114$, and an inspection of the SDSS field reveals the 
presence of a group at this redshift.  There is a nearby \redmapper\ candidate match, 
RM 11263.  The redshift of RM 11263 is just over $3\sigma$ away over the PXXIX redshift,
and it appears to have been affected by a blend with slightly lower redshift structures, which
also led to the cluster being miscentered.  We have evaluated $\lambda$ at the obvious optical
center of the cluster, setting $z=0.114$.   The change to the richness relative to RM 11293 was
insignificant, and the cluster remained an outlier in the $\lambda$--$M_{SZ}$ plane, with too
low a richness given $M_{SZ}$.  Consequently, the fact that the new richness may be somewhat affected
by blending is irrelevant: no blending would only exacerbate the problem.  We conclude that
RM 897 is the correct cluster match, and that the redshift assigned by PXXIX is incorrect.
Alternatively, the cluster could be considered an SZ projection effect.
\vspace{0.05in}

{\it Planck 1128: } The \Planck\ location is roughly half way between two rich \redmapper\ clusters,
RM 683 (Abell 1750, $\lambda=65.3$, $\zspec=0.088$, $\Delta\theta=4.2\arcmin$) and
RM 18367 ($\lambda=86.0$, $\zspec=0.559$, $\Delta\theta=3.2\arcmin$).  As such, it appears
to be a clear case of an SZ projection.  If we were to choose a single match, the better match is 
RM 18367, which is both richer and closer to the \Planck\ detection, whereas PXXIX assigned
this detection the redshift of the smaller, slightly more distant cluster RM 683.  However, either
association is acceptable as probed by the $\lambda$--$M_{SZ}$ scaling relation.
We note that there exist deep {\it XMM} images of this field which
easily detect the background cluster RM 18367.
\vspace{0.05in}

{\it Planck 1216: } This cluster is next to a bright star.  
At the reported \Planck\ location, the masked fraction of the cluster is 0.17.  \redmapper\
finds the cluster, but the assigned center (which is incorrect) is such that the masked fraction is above our mask cut of $0.2$.  
Consequently, the cluster matching procedure failed.    This cluster has an obvious starburst central galaxy,  which
led to the incorrect center in \redmapper, which in turn led to the cluster missing from the catalog.  
This is a clear form of incompleteness in the \redmapper\ catalog due centering.


\label{lastpage}

\end{document}